\documentclass{aa}
\usepackage{graphicx}
\usepackage{txfonts}
\newcommand{\ms}{\ensuremath{\mathrm{m\,s}^{-1}}}
\newcommand{\kms}{\ensuremath{\mathrm{km\,s}^{-1}}}
\newcommand{\vsi}{\ensuremath{v_\mathrm{e} \sin i}}
\newcommand{\vr}{\ensuremath{v_\mathrm{r}}}
\newcommand{\te}{\ensuremath{T_{\mathrm{eff}}}}
\newcommand{\logte}{\ensuremath{\log(T_{\mathrm{eff}})}}

\newcommand{\logl}{\ensuremath{\log(L/L_\odot)}}
\newcommand{\mmo}{\ensuremath{M/M_\odot}}

\newcommand{\bz}{\ensuremath{\langle B_z \rangle}}
\newcommand{\mo}{\ensuremath{M_\odot}}
\newcommand{\brms}{\ensuremath{B_\mathrm{rms}}}
\newcommand{\esp}{ESPaDOnS}
\newcommand{\sigb}{\ensuremath{\sigma_B}}
\newcommand{\sigbs}{\ensuremath{\sigma_B}s}

\begin{document}
   \title{Searching for links between magnetic fields and stellar evolution}
   \subtitle{III. Measurement of magnetic fields in open cluster Ap
   stars with ESPaDOnS \thanks{Based on observations made with the
   Canada-France-Hawaii Telescope, operated by the National Research 
   Council of Canada, the Centre National de Recherche Scientifique 
   of France, and the University of Hawaii, under programme 05A-C19}}
   \author{
        J.D.~Landstreet \inst{1}
       \and
        J.~Silaj        \inst{1}
       \and
        V.~Andretta     \inst{2}
       \and 
        S.~Bagnulo      \inst{3}
       \and
        S.V.~Berdyugina \inst{4,5}
       \and
        J.-F.~Donati    \inst{6}
       \and
        L. Fossati      \inst{7}
       \and
        P.~Petit        \inst{6}
       \and
        J.~Silvester    \inst{8}
       \and
        G.A.~Wade       \inst{8}
        }

   \offprints{J.~Landstreet}
\institute{Physics \& Astronomy Department,
           The University of Western Ontario,
           London, Ontario, Canada N6A 3K7. \\
           \email{jlandstr@astro.uwo.ca, jsilaj@astro.uwo.ca}
           \and
           INAF -- Osservatorio Astronomico di Capodimonte, 
           salita Moiariello 16, 80131 Napoli, Italy.\\
           \email{andretta@na.astro.it}
           \and
           Armagh Observatory,
           College Hill,
           Armagh BT61 9DG, Northern Ireland
           \email{sba@arm.ac.uk}
           \and
           Institute of Astronomy,
           ETH,
           8092 Zurich, Switzerland
           \email{sveta@astro.phys.ethz.ch}
           \and
           Tuorla Observatory, University of Turku,
           FI-21500, Piikki\"o, Finland
           \and
           Observatoire Midi-Pyr\'en\'ees,
           14, ave. Edouard-Belin,
           Toulouse, France
           \email{donati@ast.obs-mip.fr, petit@ast.obs-mip.fr}
           \and
           Institut f\"ur Astronomie, Universit\"at Wien,
           T\"urkenschanzstr. 17,
           A-1180 Wien, Austria
           \email{fossati@astro.univie.ac.at}
           \and
           Department of Physics,
           Royal Military College of Canada,
           P.O. Box 17000, Station `Forces',
           Kingston, Ontario, Canada K7K 7B4. 
           \email{Gregg.Wade@rmc.ca}
	   }

   \date{Received: September 15, 2009; accepted: ever hopeful}
%
%
  \abstract
  {A small fraction of upper main sequence stars have strong, highly
    structured magnetic fields.  The origin and evolution of these
    fields are not adequately understood.}
  {We are carrying out a survey of magnetic fields in Ap stars
    in open clusters in order to obtain the first sample of magnetic
    upper main sequence stars with precisely known ages. These data
    will constrain theories of field evolution in these stars. }
  {A survey of candidate open cluster magnetic Ap stars was carried out
    using the new ESPaDOnS spectropolarimeter at the CFHT.  This
    instrument provides an  alternative to the FORS1
    spectropolarimeter used up to now for this survey.}
  {We have obtained 44 measurements of the mean longitudinal fields
    \bz\ of 23  B6 -- A2 stars that have been identified as
    possible Ap stars and that are possible members of open clusters,
    with a median uncertainty of about 45~G. Of these stars, 10 have
    definite field detections. Nine stars  of our sample are found
      not to be magnetic Ap stars. These observations significantly
    increase the information available about low-mass stars near the
    TAMS compared to our previous sample. }
  {We find that \esp\ provides field measurements comparable to those
    that we have previously obtained with FORS1, and that these data
    also contain a large amount of useful information not readily
    obtained from lower resolution spectropolarimetry. With the new
    data we are able to expand the available data on low-mass,
    relatively evolved Ap stars, and identify more robustly which
    observed stars are actually magnetic Ap stars and cluster
    members. Re-analysis of the enlarged data set of cluster Ap stars
    indicates that such stars with masses in the range of 2 -- 5 \mo\
    show RMS fields larger than about 1~kG only when they are near the
    ZAMS. The time scale on which these large fields disappear varies
    strongly with mass, ranging from about 250~Myr for stars of 2 -- 3
    \mo\ to 15~Myr for stars of 4 -- 5 \mo. Our data are consistent
    either with emergent flux conservation for most (but not all) Ap
    stars, or with modest decline in flux with age.}


\titlerunning{A survey of open cluster Ap stars with ESPaDOnS}

\authorrunning{J.D.~Landstreet, J.~Silaj, V.~Andretta et al.}

\maketitle

\section{Introduction}

The discovery that some A- and B-type stars of the upper main sequence
possess strong (kG) magnetic fields was made more than 60 years ago
(Babcock \cite{Bab47}).  Since that time, there has been much
investigation into the occurrence and structure of these fields, but
much about them remains unknown. In particular, the evolution of these
fields with time is a major area of ignorance, and until quite
recently, almost no observational constraints have been available to
guide and test theoretical ideas about field evolution.

In the past few years the observational situation has changed
dramatically with the development of powerful new spectropolarimeters
such as the one developed for the MuSiCoS spectrograph at the TBL, and
FORS1 on the ESO VLT. With FORS1 one can obtain routine magnetic
measurements of main sequence stars fainter than $m_V \sim 10^{\rm m}$, and
these new instruments have also made possible a general decrease in
field measurement uncertainty by a factor of several.

The new generation of spectropolarimeters has for the first time made
it practical to study observationally the evolution of magnetic fields
in magnetic Ap stars with time, by exploring a large sample of such
stars that are members of open clusters. Cluster studies open the
possibility of obtaining a sample of magnetic stars with reasonably
accurate ages, thus providing a statistical time sequence of upper
main sequence magnetism.  Photometric and spectroscopic surveys have
identified a substantial sample of probable Ap stars in such clusters
which can be surveyed for fields with the new instruments, and the
membership of these stars in their clusters can be established much
more securely than in the past thanks to recent advances in
astrometry, especially the Hipparcos and Tycho-2 projects.

These new possibilities have stimulated a major survey of magnetic
fields in cluster Ap stars, whose first results have been reported by
Bagnulo, Landstreet, Mason et al. (\cite{Bagetal06}, hereafter Paper I)
and analysed by Landstreet, Bagnulo, Andretta et al. (\cite{Lanetal07},
hereafter Paper II). In the work completed so far, we have obtained
and collected magnetic measurements of 81 stars which are both
probable magnetic Ap stars and probable cluster members. The absolute
(log) ages of the stars in this sample are known with accuracy in the
range of 0.05 -- 0.2 dex (the age uncertainty of the clusters
themselves). The fraction of the main sequence lifetime elapsed (the
``fractional age'') of each of the stars of this sample is known with
a precision that is usually less than half the fractional age itself,
and thus for stars of small fractional age, the fractional age is also
quite precise. 

From this sample, we have established conclusively that fields are
present in Ap stars more massive than $2 M_\odot$ from the ZAMS
onwards. We have also shown that stars more massive than about $3
M_\odot$ have a substantial decline in field strength after the age of
about $3\,10^7$ yr, while less massive stars show no strong evidence
of change of field strength even as long as $10^9$ yr after reaching
the ZAMS.

Recently the powerful new spectropolarimeter ESPaDOnS has become
available at the CFHT. This instrument combines a large resolving
power with high throughput and wide spectral coverage, and is
the most powerful instrument for field measurement in stars with
reasonably rich spectra and low \vsi\ that is currently available. 
We have used \esp\ to extend the survey described and analysed in
Papers I and II. 

This extension of our previous survey has several aims: (1) to expand
the survey into the northern celestial hemisphere, adding clusters not
visible from ESO; (2) to increase the number of observations of stars
for which only one or two magnetic measurements are available, thereby
improving the precision of \brms\ values available for stars of the
survey; (3) to search for fields in probable magnetic Ap stars for
which no fields have yet been detected by the available observations,
exploiting the superior ability of \esp\ to detect weak fields, and to
detect fields even when the mean longitudinal field \bz\ is close to
zero (see discussion below); and (4) to compare the field measurement
accuracy obtained with high- and low-resolution spectropolarimetry for
various kinds of magnetic Ap stars.

The next section of this paper discusses the observational strategy
employed, the data reductions, and the choice of stars to observe.
Sect. 3 describes the transformation of the polarisation measurements
into measurements of \bz, and tests of the data quality. Sect. 4
presents the new field measurements obtained. In this section we also
compare the observational efficiencies of various instruments. Sect. 5
contains a star-by-star assessment of the new information obtained
from the \esp\ observations. In Sect. 6 we discuss general features
of the new addition to our data set of cluster magnetic Ap stars. In
Sect. 7 the new cluster field measurements are integrated into our
larger survey, and the conclusions of Paper II are reviewed in the
light of our new data.  Finally Sect. 8 summarises the conclusions of
the paper.

\section{Instrumentation, observations and data reduction}

\subsection{\esp} 

The instrument used for this portion of our survey of magnetic Ap
stars in clusters is \esp, the new cross-dispersed echelle
spectropolarimeter built for the Canada-France-Hawaii telescope. The
instrument is conceptually similar to the MuSiCoS spectropolarimeter
which has been extensively used for high-precision magnetic
measurements (e.g. Wade et al. \cite{Wadetal00a}, \cite{Wadetal00b}),
but has a factor of about 30 times higher efficiency.

The polarisation analyser is located at the Cassegrain focus of the
telescope. The stellar image is formed on an aperture followed by a
collimating lens. The beam then passes through a rotatable $\lambda/2$
waveplate, a fixed $\lambda/4$ waveplate, a second rotatable
$\lambda/2$ waveplate, and finally a small-angle Wollaston prism,
followed by a lens which refocuses the (now double) star image on the
input of two optical fibres. This relatively complex polarisation
analyser is necessary because one of the fundamental design parameters
for \esp\ was very wide wavelength coverage (approximately 3700
\AA\ to 1.04 $\mu$m). To have waveplates which are approximately
achromatic over this wide range, \esp\ uses Fresnel rhombs. A single
Fresnel rhomb acts as a $\lambda/4$ plate, but deviates the beam,
while two Fresnel rhombs in series form a $\lambda/2$ plate without
beam deviation. To minimise mechanical complications, only the double
(non-deviating) Fresnel rhombs are allowed to rotate; the
configuration chosen is the minimum which allows one to analyse all of
the Stokes polarisation components $(Q, U, V)$ by appropriate
orientation of the axes of the successive waveplates.

The two output beams from the Wollaston prism, which have been
analysed into the two components of linear or circular polarisation as
desired by appropriate waveplate settings, are then carried by the
pair of optical fibres to a stationary and thermally buffered (but not
yet temperature-controlled) cross-dispersed spectrograph where two
interleaved spectra are formed, covering virtually the entire desired
wavelength range with a resolving power of $R \approx 65\,000$. The
$I$ component of the stellar Stokes vector is formed by adding the two
corresponding spectra, while the desired polarisation component $(Q,
U, V)$ is obtained essentially from the difference of the two
spectra. To minimise the systematic errors due to small misalignments,
differences in transmission, effects of seeing, etc., one complete
polarisation observation of a star consists of four successive spectra
taken with four different positions of the two rotatable Fresnel
rhombs; for the second and third exposures, the waveplate settings
exchange the right and left circularly analysed beams with respect to
the handedness measured in the first and fourth exposures (cf. Donati
et al. \cite{Donetal97};  Donati et al. in preparation;
  http:www.cfht.hawaii.edu/Instruments/Spectroscopy/Espadons/).

\subsection{Observations} 

The stars observed were selected from a large data-base we have
assembled (see Papers I and II) of possible magnetic Ap stars that are
also possible members of open clusters within about 1 kpc of the Sun.
As discussed in Paper II, we have used available astrometric and
photometric data, especially those from the Hipparcos (ESA
\cite{Esa97}) and Tycho-2 (H\o g et al. \cite{Hogetal00a},
\cite{Hogetal00b}) catalogues, to clarify as far as possible which
magnetic Ap stars are probable or definite cluster members. Here we
use essentially the same criteria for cluster membership that were
discussed in Paper II. We selected stars from the data-base that were
observable in July 2005, when these observations were carried out. In
part, we observed stars that are not observable from ESO, where we
have obtained the previous observations for this survey.  However, a
number of observations were obtained of stars for which previous
measurements are available, either by us or by others.

As discussed in detail in Paper II, the classification of most of the
stars observed as magnetic Ap stars is still fairly uncertain, because
identification of cluster Ap stars is usually based only on narrow-band
photometry and/or low dispersion, low signal-to-noise spectra. Thus we
expected that some of the stars observed would turn out not to be
magnetic Ap stars at all. 

Our previous observations for this survey were carried out using the
low-resolution spectropolarimeter FORS1 at the ESO VLT. With $R \sim
10^3$, FORS1 magnetic field measurements give high weight to the
circular polarisation signal in the Balmer lines, which are resolved
even at this low dispersion. In cases where a relatively rich metallic
spectrum is present, the FORS1 measurements can be made significantly
more precise by exploiting the polarisation signal present in the
(barely resolved) metallic line spectrum, but to first order the
precision of FORS1 magnetic field measurements is not sensitive to the
details of the metallic line spectrum. In contrast, measurements with
\esp, like those obtained with MuSiCoS, have high enough resolving
power to fully exploit the information content of the metallic line
spectrum. \esp\ field measurements have standard errors which
decrease strongly with decreasing \vsi\ and with increasing richness
and strength of the metallic line spectrum (cf. Landstreet
\cite{Lan82}; Shorlin et al. \cite{Shoetal02}). To fully exploit this
dependence, and thus to obtain the most precise measurements possible,
we have preferentially observed relatively cool magnetic Ap
stars, and have selected our targets for low \vsi\ where available
data allowed us to select targets with this criterion. 

The list of stars observed is given in Table~\ref{stars-obs.tab},
which contains the name of each star, the cluster to which it may
belong, our assessment of the probability that the star is actually a
cluster member (Y = definite member, P = probable member, ? =
questionable, assessed as in Paper II), the spectral
type based on a combination of published types with the results of our
own examination of individual $I$ spectra, and the \vsi\ value deduced
from our \esp\ spectra (the uncertainty in these values is
approximately the larger of $\pm 10$\% or $\pm 2$~\kms).

\begin{table}
\caption{\label{stars-obs.tab} Stars observed with \esp\ for magnetic fields } 
\begin{tabular}{rrrrrrrrrr} 
\hline 
\hline \\ 
  Star &  Cluster &  Memb &   Spectrum  &  $v \sin i$  \\ 
   &   &   &   &  (\kms) \\ 
\hline \\ 
   HD 16605  &       NGC 1039  &    P  &   A1p SiCrSr   &    18  \\ 
   HD 16728  &       NGC 1039  &    Y  &         B9V    &    75  \\ 
   HD 19805  &   $\alpha$ Per  &    Y  &       B9.5V    &     8  \\ 
  HD 108945  &       Coma Ber  &    Y  &     A2p SrCr   &    65  \\ 
  HD 144661  &      Upper Sco  &    Y  &    B8p He-wk   &    45  \\ 
  HD 153948  &       NGC 6281  &    P  &        Ap Si   &    80  \\ 
  HD 317857  &       NGC 6383  &    ?  &       A1IVp    &    26  \\ 
  HD 318100  &       NGC 6405  &    Y  &         B9p    &    42  \\ 
  HD 320764  &       NGC 6475  &    P  &       A1V an   &   225  \\ 
  HD 162576  &       NGC 6475  &    Y  &   B9.5p SiCr   &    30  \\ 
  HD 162588  &       NGC 6475  &    Y  &        Ap Cr   &    73  \\ 
  HD 162630  &       NGC 6475  &    P  &         B9V    &    51  \\ 
  HD 162656  &       NGC 6475  &    Y  &         B9V    &    45  \\ 
  HD 162725  &       NGC 6475  &    Y  &     A0p SiCr   &    32  \\ 
  HD 169842  &       NGC 6633  &    Y  &     A1p SrCr   &    45  \\ 
 HD 169959A  &       NGC 6633  &    ?  &       A0p Si   &    57  \\ 
  HD 170054  &       NGC 6633  &    Y  &        B6IV    &    26  \\ 
  HD 170860  &        IC 4725  &    P  &      B9IV/Vp   &    85  \\ 
  HD 172271  &        IC 4756  &    P  &       A0p Cr   &   107  \\ 
  HD 205073  &       NGC 7092  &    Y  &          A1    &    16  \\ 
  HD 205331  &       NGC 7092  &    Y  &         A1V    &    51  \\ 
 BD+49 3789  &       NGC 7243  &    P  &       B7p Si   &    81  \\ 
 HIP 109911  &       NGC 7243  &    Y  &      A0Vp Si   &    59  \\ 
\hline 
\hline 
\end{tabular}

\end{table}

\subsection{Data Reduction}

The observed polarised spectra were reduced to 1D using the dedicated
software LibreEsprit, provided by J.-F. Donati for treatment of \esp\
data. LibreEsprit subtracts bias, locates the various spectral orders
on the CCD image, measures the shape of each order and models the
(varying) slit geometry, identifies comparison lines for each order
and computes a global wavelength model of all orders, performs an
optimal extraction of each order, and combines the resulting spectra
(in groups of four, corresponding to the four sub-observations with
the four different Fresnel rhomb settings) to obtain intensity ($I$)
and circular polarisation ($V$) spectra. The $V$ spectrum normally has
the continuum polarisation removed, as this arises mainly from
instrumental effects and carries little information about the
star. Each spectrum is corrected to the heliocentric frame of
reference, and may optionally be divided by a flat field and be
approximately normalised (see Donati et al.  \cite{Donetal97}, and Web
pages at
http://www.cfht.hawaii.edu/Instruments/Spectroscopy/Espadons).

A check spectrum called the $N$ spectrum, computed by combining the
four observations of polarisation in such a way as to have real
polarisation cancel out, is also calculated by LibreEsprit. The $N$
spectrum tests the system for spurious polarisation signals. In all of
our observations, the $N$ spectrum is quite featureless, as expected. 

The final spectra consist of ASCII files tabulating $I$, $V$, $N$, and
estimated uncertainty per pixel as a function of wavelength, order by
order. We have kept both normalised and un-normalised versions of each
spectrum.

\section{Measurement of magnetic field strength}

Two methods allow us to detect magnetic fields in the stars
observed. First, we can use the $V$ spectra to measure the mean
longitudinal magnetic field strength \bz\ of each star at the time of
observation. This is the conventional measure of field strength
normally used for detection of fields in magnetic Ap stars
(e.g. Landstreet \cite{Lan82}). However, because of the high value of
resolving power, we can also examine spectral lines for the presence
of circular polarisation signatures; Zeeman splitting combined with
Doppler broadening of lines by rotation can lead to non-zero values of
$V$ within spectral lines even when the value of \bz\ is close to
zero. This possibility substantially increases the sensitivity of our
measurements as a discriminant of whether a star is in fact a magnetic
star or not, as discussed by Wade et al. (\cite{Wadetal00a}) and
Shorlin et al. (\cite{Shoetal02}).

Because \esp\ is a high-resolution instrument and because the stars
observed are relatively faint (typical magnitudes $m_V$ between 6 and
10), the signal-to-noise ratio (SNR) in individual spectral lines is
not very high, and field measurements made with single lines are not
very precise. In this situation, we employed the technique of Least
Squares Deconvolution (LSD), developed into a practical strategy by
Donati et al. (\cite{Donetal97}) and employed extensively in the
analysis of magnetic observations made with the MuSiCoS
spectropolarimeter (cf Wade et al. \cite{Wadetal00a}).

In the LSD technique, we start with a list of spectral lines expected
in a normal or a magnetic Ap star of a temperature near that of the
star observed. This line list is obtained from the Vienna Atomic Line
Database (VALD; Piskunov et al. \cite{Pisetal95}; Ryabchikova et
al. \cite{Ryaetal99}; Kupka et al. \cite{Kupetal99}), and is selected by
specifying a suitable chemical composition, wavelength range, and
minimum line depth. For the stars in our sample, two different
abundance tables were employed.  An Ap abundance table was created by
assigning the iron-peak elements abundances $10 \times$ solar
abundance, except for chromium which was increased to $100 \times$
solar. A solar abundance table is the default provided by VALD when a
request is made, and was used to make the remaining lists. VALD
line lists were requested to include all lines in the ESPaDOnS spectral
range (3700-10000~\AA) with depths greater than 10\% of the continuum
(in a synthetic spectrum with $\vsi = 0$).

Fundamental parameters (effective temperature \te\ and surface
gravity) for the stars observed were determined primarily from
Str\"{o}mgren and/or Geneva photometry, as described in detail in Paper
II.  Neither Str\"{o}mgren photometry nor Geneva photometry is available
for the stars HD~169842, HD~172271 and HD~317857, so the effective
temperatures were estimated from Johnson UBV photometry or the
observed spectra. Initial selection of an appropriate LSD line list
for each star was made based on these values. Experimentation showed
that the highest SNR in the measured field \bz\ generally resulted
from the use of a line list selected with a temperature close to, or
up to about $10 - 15$\% hotter than, the estimated \te\ of the star.
 (Using a line list appropriate for a slightly hotter star seems to
eliminate a number of weak lines that contribute more to the noise
than to the signal.)

For each spectral line in the line list appropriate to a particular
star, a small window is cut out of the observed $I$ and $V$ spectra
centred on the laboratory wavelength of the line, and the local wavelength
scale is converted to a velocity scale relative to the nominal central
wavelength. Conceptually, one may imagine that all the small $I$ and
$V$ spectral windows are then averaged. (In practice, model intensity and
polarisation signatures are fit to the ensemble of single line data
windows, rather than performing direct averaging.) In this average,
each window is given a weight proportional to the line depth, the
wavelength, and the Land\'e factor of the line. Since the $I$ and $V$
signatures in individual spectral lines are very similar to one
another, and scale essentially with the weighting factors used in
averaging, the resulting mean intensity and polarisation line profiles
may be analysed as if they represented a single real spectral line
having a much higher SNR than the individual observed lines. The
improvement in SNR from single lines to the LSD line is by roughly the
square root of the number of lines used, so the SNR in the LSD mean
line is typically of order 30 times higher than in single lines.

The LSD line is then analysed in two ways. First, the value of \bz\ is
computed using the normal expression (e.g. Donati et al. \cite{Donetal97})
\begin{equation}
\langle B_{\rm z} \rangle = -2.14\,10^{12} \frac{\int v V(v) dv}
              {\lambda z c \int [I_{\rm c} - I(v)] dv},
\end{equation}
where \bz\ is in G, $z$ is the mean Land\'e factor of the LSD line
(typically about 1.25), and $\lambda$ is the weighted mean wavelength
of the LSD line in \AA\ (typically about 5200 \AA).  The limits of
integration were chosen visually for each star to coincide with the
apparent edges of the LSD $I$ and $V$ lines; using a smaller window
would neglect some of the signal coming from the limb of the star,
while a window larger than the actual line would increase the noise
without adding any further signal, thus degrading the SNR below the
optimum value achievable. If the resulting value of \bz\ is
significantly non-zero (say by $4 - 5 \sigb$ in a single observation,
or by $3 \sigb$ in two or more observations), we conclude that a
magnetic field has been detected.

In addition, the LSD line profile itself is examined. If the value of
$V$ rises to a statistically significant value inside the line, while
always remaining insignificant in the neighboring continuum, we
conclude that a field is detected. As mentioned above, this may occur
even if \bz\ is not significantly different from zero. On the other
hand, from our experience with searches for weak fields in known Ap
stars (cf Auri\`ere et al. \cite{Auretal07}), a star with a favourable
spectrum (hundreds of lines and \vsi\ less than about 50 or 60 \kms)
which shows no significantly non-zero values of $V$ within the LSD
spectral line in two independent high SNR observations is unlikely to
host a magnetic field. Furthermore, since observations with \esp\
yield high-resolution $I$ spectra (as well as $V$ spectra), we may
test the suspicion that a star is in fact not a magnetic Ap star at
all by examining the $I$ spectrum. The combination of no significant
$V$ signature with a normal $I$ spectrum has confirmed that several of
the stars selected for observation are indeed not Ap stars.

\begin{figure}
\resizebox{9.0cm}{!}{
\includegraphics*[angle=0]{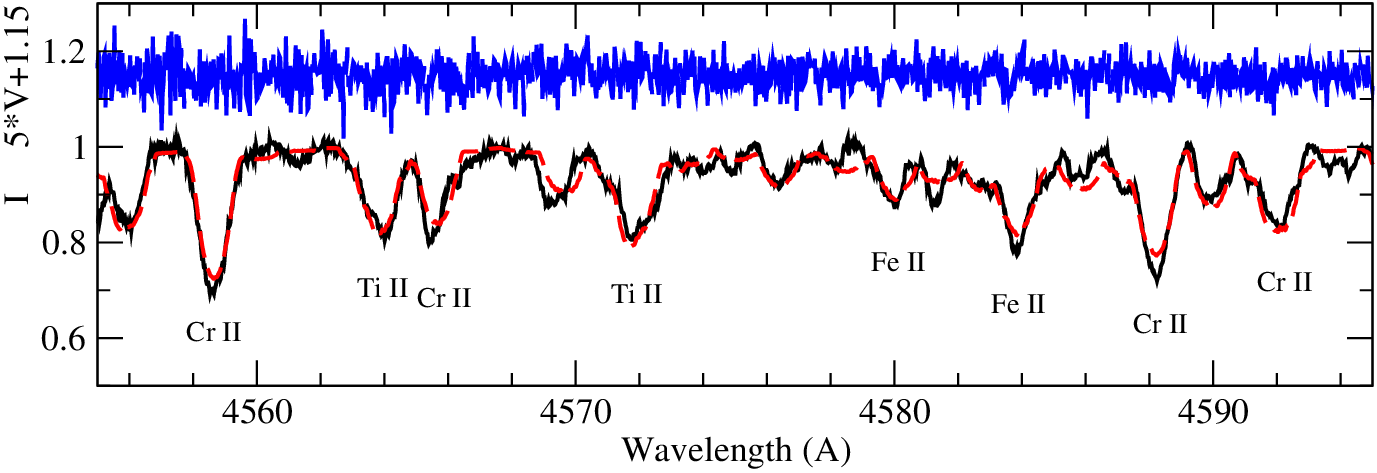}}
\caption{\label{hd169842-458} A sample region of the $I$ spectrum of
HD~169842 with identification of strong lines (lower solid line) is
compared to a synthetic spectrum (dashed line), for which Cr has been
enhanced by a factor of about 80. Above is the same
region of the $V$ spectrum, multiplied by 5 and displaced upward by
1.15 for clarity. 
}  
\end{figure}

\begin{figure} \resizebox{9.0cm}{!}{
\includegraphics*[angle=0]{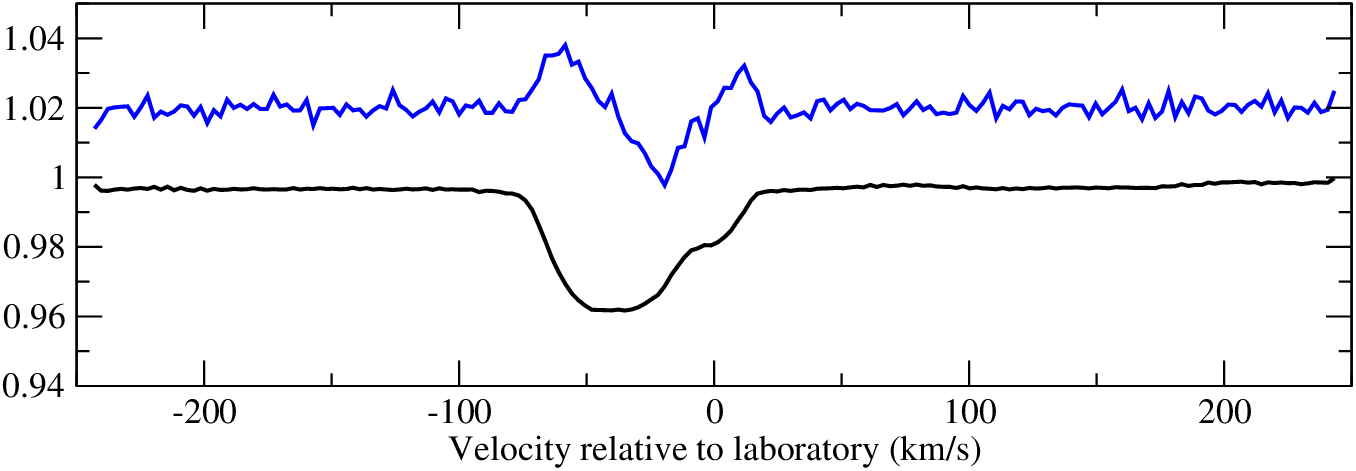}}
\caption{\label{hd169842-lsd} 
The LSD $I$ (lower) and $V$ (upper; multiplied by 25 and displaced
upwards by 1.02 for visibility) spectra of HD~169842 from the same
night as Figure~\ref{hd169842-458}.  
}
\end{figure}

As an example, small segments of the \esp\ $I$ and $V$ spectra of
HD~169842 from JD~2453566.902 are shown in Figure~\ref{hd169842-458}.
The $I$ spectrum is compared to a synthetic spectrum computed using
the magnetic line synthesis programme Zeeman (cf. Landstreet
\cite{Lan88}) with (nearly) solar abundances except for Cr, which is
enhanced by 1.9 dex. Note that no trace of Zeeman polarisation is seen
in individual spectral lines; in this star the Zeeman signal is too
small relative to the noise level in $V$ to be directly visible. The
LSD $I$ and $V$ spectra computed from these data are shown in
Figure~\ref{hd169842-lsd}. Averaging over 4836 spectral lines has
reduced the noise in the LSD $V$ spectrum by a factor of order 30
relative to the noise level in individual spectral lines, allowing the
magnetic field signature in $V$ to be clearly seen with an amplitude
well above the small noise level outside the line.  (The number of
lines used in the LSD mask is a strong function of both \te\ and
degree of peculiarity. The Ap masks used have about 8300 lines for a
star of $\te \sim 8500$~K, decreasing smoothly to about 1600 lines at
$\te \sim 15000$~K. Typically the masks for normal composition have
about 1/3 as many lines as the Ap masks.)

\subsection{Verification}

One important kind of verification of the performance of \esp\ is to
confirm that it does not produce spurious magnetic field
detections. We have tested this in two ways. First, we have made a
precise field measurement ($-9 \pm 19$ G) of the apparently
non-magnetic star $\alpha$ And (= HD~358), which, in agreement with
numerous other measurements made with a variety of instruments, shows
no significant field (Wade et al. \cite{Wadetal06}). Secondly, we find
in the present data a number of stars which turn out (on the basis of
our new high-resolution $I$ spectra) to be normal rather than magnetic
Ap stars; invariably such stars also do not show any significant
magnetic field.

\begin{figure}
\resizebox{8.0cm}{!}{
\includegraphics*[angle=0]{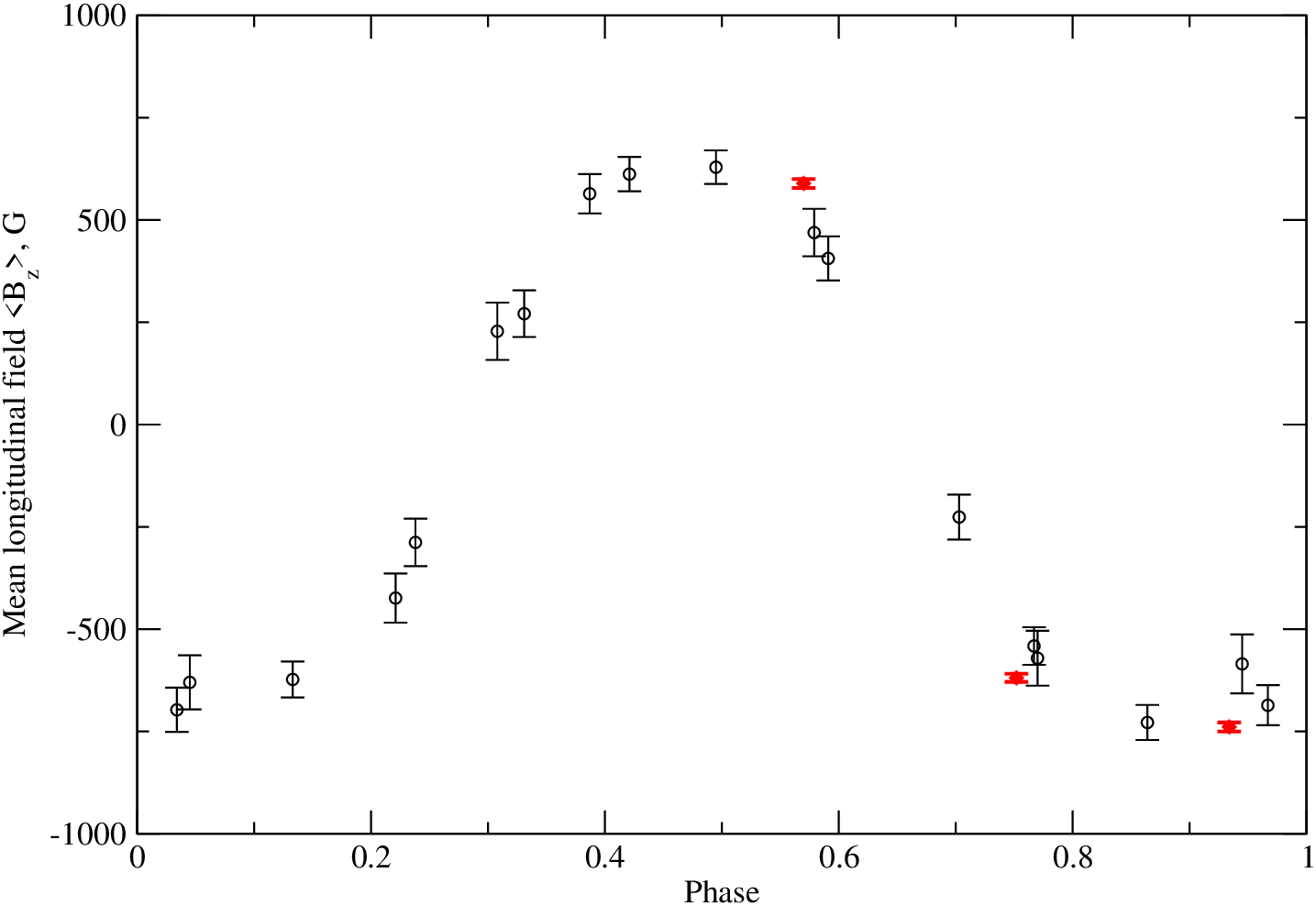}}
\caption{\label{alp2cvn-comp.fig} Phased magnetic mean longitudinal
  field variations of $\alpha^2$ CVn as reported by Wade et
  al. \cite{Wadetal00b} (open circles). Our three new data points
  (filled circles) are those with very small error bars. 
}
\end{figure}

We have also verified that our field measurements are consistent with
earlier work by observing the longitudinal field of the well-studied
magnetic Ap star $\alpha^2$ CVn during three nights. These data are
presented in Table~\ref{alp2cvn.tab}. We have computed phases (at the
mid-point of each observation) according to the ephemeris of Wade et
al. (\cite{Wadetal00b}), and we compare the three new measurements
with their magnetic curve in Figure~\ref{alp2cvn-comp.fig}. It is
clear from the figure that the new observations are fully consistent
with the previous (MuSiCoS) data obtained and reduced in a very
similar way, but the ESPaDOnS data are about a factor of five more
precise.

\begin{table}[h!] 
\caption{\label{alp2cvn.tab} Observations of standard star 
   $\alpha^2$ CVn}
\begin{tabular}{rrrrrr}
\hline
\hline
\\
HJD           &  SNR  &  phase  &  $\langle B_{\rm z} \rangle \pm 
   \sigb$ \\ 
              &       &         &   (G)                            \\
\hline
2453569.748   & 680   & 0.570 $\pm 0.01$  &   589 $\pm$ 11 \\
2453570.747   & 1040  & 0.752 $\pm 0.01$  &  $-619 \pm$ 10 \\
2453571.738   & 580   & 0.934 $\pm 0.01$  &  $-739 \pm$ 11 \\
\hline 
\hline 
\end{tabular}
\end{table}

\begin{table*}[h!] 
\caption{\label{fld-meas.tab} Magnetic field measurements of possible cluster Ap stars } 
\begin{tabular}{rrrrrrrrrr} 
\hline 
\hline \\ 
        Star & HJD & SNR & $v_{r}$ & $\langle B_{\rm z} \rangle \pm \sigma_{\rm B}$ & fld \\ 
  &  &  &  (\kms)  & (G)  \\ 
\hline \\ 
    HD 16605 &   2453570.110  &   608  &    -1  &   -2099  $\pm$   34 & DD \\ 
             &   2453571.081  &   689  &    -1  &   -2205  $\pm$   29 & DD \\ 
    HD 16728 &   2453572.123  &  1510  &        &      -2  $\pm$  272 &    \\ 
    HD 19805 &   2453570.131  &  1045  &     0  &       6  $\pm$   24 &    \\ 
             &   2453572.129  &  1453  &     0  &      -3  $\pm$   14 &    \\ 
   HD 108945 &   2453569.732  &  1884  &    +1  &    -111  $\pm$   28 & DD \\ 
             &   2453569.740  &  1872  &    +1  &     -75  $\pm$   29 & DD \\ 
             &   2453570.735  &  2487  &    +1  &     109  $\pm$   20 & DD \\ 
   HD 144661 &   2453571.758  &  2814  &    -3  &     -55  $\pm$   48 &    \\ 
             &   2453571.758  &  2946  &    -4  &      11  $\pm$   48 &    \\ 
   HD 153948 &   2453569.768  &  1005  &    -2  &    -222  $\pm$   82 & DD \\ 
             &   2453570.774  &   942  &    -2  &    -185  $\pm$   89 & DD \\ 
             &   2453571.778  &   901  &    -3  &     199  $\pm$   90 &    \\ 
   HD 317857 &   2453571.810  &   568  &   -10  &    -919  $\pm$   20 & DD \\ 
   HD 318100 &   2453569.804  &   597  &    -8  &    -644  $\pm$   59 & DD \\ 
   HD 162576 &   2453568.870  &  1861  &   -14  &      10  $\pm$   25 & DD \\ 
             &   2453569.828  &  1568  &   -13  &     -19  $\pm$   28 &    \\ 
             &   2453570.797  &  2498  &   -13  &      13  $\pm$   18 &    \\ 
   HD 162588 &   2453568.884  &  1625  &   -16  &      93  $\pm$   84 &    \\ 
             &   2453569.846  &  1039  &   -15  &      41  $\pm$  141 &    \\ 
             &   2453571.858  &  2016  &   -13  &      36  $\pm$   70 &    \\ 
   HD 162630 &   2453570.819  &  1229  &   -15  &     -61  $\pm$  117 &    \\ 
             &   2453571.872  &  1407  &   -13  &    -157  $\pm$  116 &    \\ 
   HD 162656 &   2453571.885  &  1338  &    +3  &      24  $\pm$   78 &    \\ 
   HD 162725 &   2453566.876  &  1654  &   -14  &    -101  $\pm$   14 & DD \\ 
             &   2453569.865  &  1608  &   -13  &     -21  $\pm$   20 & DD \\ 
   HD 320764 &   2453568.858  &  1212  &        &    -399  $\pm$  408 &    \\ 
             &   2453571.832  &  1131  &        &     137  $\pm$  448 &    \\ 
   HD 169842 &   2453566.902  &  1395  &   -30  &     180  $\pm$   20 & DD \\ 
             &   2453569.890  &   930  &   -30  &     230  $\pm$   30 & DD \\ 
             &   2453570.858  &  1441  &   -30  &    -153  $\pm$   21 & DD \\ 
  HD 169959A &   2453570.885  &  1378  &   -53  &    -691  $\pm$  113 & DD \\ 
   HD 170054 &   2453569.925  &   781  &   -28  &      12  $\pm$  108 &    \\ 
             &   2453570.888  &  1556  &   -26  &     -87  $\pm$   54 &    \\ 
   HD 170860 &   2453571.901  &  1005  &   -13  &     462  $\pm$  236 & DD \\ 
   HD 172271 &   2453567.897  &  1441  &   -21  &    -265  $\pm$   90 &    \\ 
   HD 205073 &   2453567.914  &  1395  &    -7  &       4  $\pm$   15 &    \\ 
             &   2453569.972  &  1102  &   +18  &      -8  $\pm$   21 &    \\ 
   HD 205331 &   2453567.922  &  1614  &    -5  &     -49  $\pm$   41 &    \\ 
             &   2453570.003  &  2022  &    -4  &      -9  $\pm$   33 &    \\ 
             &   2453570.967  &  1688  &    -3  &     143  $\pm$   41 &    \\ 
 BD +49-3789 &   2453570.941  &   528  &    -6  &     561  $\pm$  400 &    \\ 
  HIP 109911 &   2453570.033  &   545  &   -10  &     494  $\pm$  230 &    \\ 
\hline 
\hline 
\end{tabular}

\end{table*}

\section{Results}

The field measurements obtained during the \esp\ survey are presented
in Table \ref{fld-meas.tab}. This table lists the stars for which
field measurement was successful; the heliocentric Julian date (HJD)
of each observation; the signal-to-noise ratio (SNR) achieved per \AA\
of spectrum, derived from the value reported in the output file of
LibreEsprit for the $N$ spectrum of order 45 at 5030 \AA; the measured
value of the stellar heliocentric radial velocity \vr\ (in \kms; generally
accurate to about $\pm 1 - 2$~\kms); the measured mean longitudinal
field strength \bz\ (in G) with its standard error (computed by treating each
integration pixel as independent, using the uncertainty per pixel of
$V$ assigned by the LSD process); and finally a flag (fld) that indicates
whether a field has been definitely detected (DD) or not (blank). Note
that the radial velocity has been measured from the LSD spectrum by
bisecting the line wings close to where they reach the continuum. The
precision of such a measurement is limited primarily by the width and
shallowness of the line profile and occasionally by the unusual shape;
the absolute precision of \esp\ radial velocities for very sharp-line
stars is a few hundred \ms, and the stability is on the order of a few
tens of \ms.

As discussed above, a field may be detected through the $V$ signature
even if \bz\ is close to zero, as is the case for example of one
observation each of HD~108945, HD~162576 and HD~162725. On the other
hand, the field of a star which is actually a magnetic Ap star may not
be detected, for example if the field is particularly weak (of order
100~G or less), if the spectrum is strongly broadened by rotation, if
\te\ is high and there are relatively few metal lines, or if the SNR
of the spectrum obtained is too low.

It is of interest to compare the relative measurement efficiencies of
magnetic measurements made with low and high spectral resolution. We
can obtain such a comparison by examining measurements of the same
star made with FORS1 and \esp. Fortunately we have in our data sets
several stars observed with both instruments.

\begin{table*}
\begin{center}
\caption{\label{esp-fors-comp.tab} Comparison of ESPaDOnS and FORS1
  measurements of the same stars}
\begin{tabular}{rrrrrrrrrr}
\hline
\hline \\
Star       & $m_V$ & \te  & \vsi & Instrument & HJD        & $\Delta t$ & SNR &
       $\bz \pm \sigma_{\rm B}$ & fld  \\
           &       & (K)  & (\kms) &          &            & (min)      &     &
           (G)   \\
\hline \\
HD 108945  & 5.46  &  8800 & 65  & FORS1    &              & 15 & 1250 & $-44  \pm$ 75     & nnn \\
           &       &       &     & ESPaDOnS & 2453569.732  & 12 & 1880 & $-111 \pm$ 28     & DD  \\
           &       &       &     & ESPaDOnS & 2453569.740  & 12 & 1870 &  $-75 \pm$ 29     & DD  \\
           &       &       &     & ESPaDOnS & 2453570.735  & 15 & 2490 &  109 $\pm$ 20     & DD  \\
HD 153948  & 9.35  & 10600 & 80  & FORS1    &              & 70 & 3100 &  195 $\pm$ 53     & DnD \\
           &       &       &     & ESPaDOnS & 2453569.768  & 40 & 1000 & $-222 \pm$ 82     & DD  \\
           &       &       &     & ESPaDOnS & 2453570.774  & 38 &  940 & $-185 \pm$ 89     & DD  \\
           &       &       &     & ESPaDOnS & 2453571.778  & 37 &  900 &  199 $\pm$ 90     & ND  \\
HD 317857  & 10.30 & 10000: & 26 & FORS1    &              & 75 & 2150 & $-1558 \pm$ 54    & DDD \\
           &       &       &     & ESPaDOnS & 2453571.810  & 46 &  570 & $-919 \pm$ 20     & DD  \\
HD 318100  & 9.84  & 10600 & 42  & FORS1    &              & 47 & 2100 &  345 $\pm$ 64     & DnD \\
           &       &       &     & ESPaDOnS & 2453569.804  & 50 &  600 & $-644 \pm$ 59     & DD  \\
HD 162725  & 6.42  &  9600 & 32  & FORS1    &              & 46 & 4450 &  $-60 \pm$ 33     & nnn \\
           &       &       &     & ESPaDOnS & 2453566.876  & 20 & 1650 & $-101 \pm$ 14     & DD  \\
           &       &       &     & ESPaDOnS & 2453569.865  & 26 & 1610 &  $-21 \pm$ 20     & DD  \\
HD 320764  & 8.90  &  8500 & 225 & FORS1    &              & 36 & 2400 &  $-69 \pm$ 66     & nnn \\
           &       &       &     & ESPaDOnS & 2453568.858  & 33 & 1212 & $-399 \pm$ 408    & ND  \\
           &       &       &     & ESPaDOnS & 2453571.832  & 38 & 1131 &  137 $\pm$ 448    & ND  \\
HD 169959A & 7.57  & 11000 & 57  & FORS1    &              & 30 & 2550 & $-486 \pm$ 52     & DnD \\
           &       &       &     & ESPaDOnS & 2453570.885  & 25 & 1378 & $-691 \pm$ 113    & DD  \\
HD 170054  & 8.18  & 14500 & 26  & FORS1    &              & 42 & 2300 &   64 $\pm$ 68     & nnn \\
           &       &       &     & FORS1    &              & 36 & 2200 &   84 $\pm$ 77     & nnn \\
           &       &       &     & ESPaDOnS & 2453569.925  & 52 &  780 &   12 $\pm$ 108    & ND  \\
           &       &       &     & ESPaDOnS & 2453570.888  & 27 & 1560 &  $-87 \pm$ 54     & ND  \\
HD 170860  & 9.39  & 13700 & 85  & FORS1    &              & 45 & 2050 &  $-41 \pm$ 61     & nnn \\
           &       &       &     & ESPaDOnS & 2453571.901  & 43 & 1005 &  462 $\pm$ 236    & DD  \\
\hline
\hline
\end{tabular}
\end{center}
\end{table*}

In Table \ref{esp-fors-comp.tab} we compare measurements of the same
star made with FORS1 and with \esp. This table lists individual stars,
magnitude $m_V$, the important parameters \te\ (determined as in Paper
II) and \vsi\ which have a strong influence on the precision of \esp\
measurements, and then for specific measurements with FORS1 or \esp\
we list HJD of observation, the total time required for an observation
(including telescope setting time) $\Delta t$, the resulting field
measurement with its uncertainty, and whether a field was detected
(for FORS1,  as in Paper I, nnn means no detection, DnD means
definite detection only in Balmer lines, and DDD means detection in
both Balmer lines and the metal spectrum; for \esp\ measurements, ND
means no detection, DD means definite detection).

Note that overheads for a single magnetic field measurement with FORS1
total about 20 min (in ``fast'' mode). For stars brighter than $m_V =
8$ or 9, the integration time is small compared to the overheads, so
the time-on-target for such bright stars will be independent of
magnitude. (The justification for using FORS1 in this inefficient mode
was its uniqueness in the southern hemisphere. FORS1 showed its full
capabilities in the detection of a field in a broad-lined Ap star of
$m_V = 12.9$; see Bagnulo et al. \cite{Bagetal04}.) The overheads for a
full \esp\ magnetic measurement total about 9 min, so the
time-on-target for \esp\ exposures is dominated by the actual
exposure time for stars  fainter than about $m_V = 6$.

In examining Table \ref{esp-fors-comp.tab}, it is worth recalling that
because low-resolution spectropolarimetry, such as the FORS1
measurements, relies heavily on the polarisation in the wings of the
Balmer lines, which do not vary greatly over the spectral range of our
survey, the measurement uncertainties obtained using this class of
techniques tend to fall within a moderate range. Inspection of Table
A.3 of Paper I shows that most of the measurements there have standard
errors between about 40 and 90~G, while the most accurate measurements
have $\sigb \approx 30$~G.This is also true of the subset of data in
Table~\ref{esp-fors-comp.tab}, where all the uncertainties are in the
range of 33 -- 77~G.  In contrast, the uncertainties of
high-resolution spectropolarimetry, such as the \esp\ data, are
significantly more scattered. The precision of a high-resolution
Zeeman measurement depends strongly on the number and strength of
spectral lines, on \vsi, and on the total signal collected. Very
precise measurements (\sigb\ as low as a few G) are possible for stars
having many sharp and deep lines (compare Wade et
al. \cite{Wadetal06}). More than 1/3 of the \esp\ measurements have
$\sigb \leq 30$~G, while about 20\% of our observations have $\sigb >
100$~G; the largest \sigbs\ in Table~\ref{fld-meas.tab} are over
400~G.

Examination of Table~\ref{esp-fors-comp.tab} reveals that in similar
$\Delta t$s, \esp\ and FORS1 often obtain measurements with roughly
similar field strength uncertainties. The \esp\ measurements are
substantially less precise primarily for stars with values of \vsi\
above 60 or 70~\kms (HD~153948), for stars with \te\ above perhaps
12\,000~K (HD~170054), or for stars for which insufficient signal was
obtained due to faintness, weather or other conditions. On the other
hand, in cases of particularly favourable circumstances (cool stars
with low \vsi), the \sigbs\ can be very low, even below 20~G
(HD~19805, HD~162725, HD~205073).

The reason that an instrument such as \esp, on a 3.6-m telescope, can
be competitive with one on an 8-m telescope is partly a consequence of
the way in which the high-resolution instrument can exploit more fully
the information content of the metallic-line spectrum of the star
observed. However, other factors are also involved. In the particular
case of FORS1, observations of bright stars ($m_V < 10$) have a very
low duty cycle. Most of the observing time is spent setting on target
and closing the loop on the adaptive optics, and on reading out the
CCD. The actual shutter time on such a star is only a few minutes
out of the roughly 30 min needed for a typical bright star
observation. For stars fainter than $10^{\rm m}$, for which longer
integration times are required (e.g. 2 hr for the $m_V \approx 13^{\rm
m}$ star NGC~2244-334; Bagnulo et al. \cite{Bagetal04}), the
instrument on the larger telescope achieves a much higher duty cycle,
and the aperture advantage is much more important.

\section{Individual stars}

Because each \esp\ observation provides not only a magnetic
measurement but also a high-dispersion spectrum covering some
7000~\AA\ through the visible and near-IR, these data can provide a
number of useful pieces of information. We obtain a measurement of the
mean longitudinal field as well as supplementary information about the
presence or absence of a magnetic field from the behaviour of $V$
within the LSD $I$ line profile. We can also measure the projected
rotation velocity \vsi, and the stellar radial velocity \vr. The value
of \vsi\ provides a constraint on possible rotation periods, and \vr\
provides a potentially useful test of cluster membership. We can use
the shape of the LSD $I$ profile, and its possible variations if we
have multiple observations, as valuable additional clues to test
whether a doubtful star is actually a magnetic Ap star. Finally, we
can examine (and even model) the observed $I$ spectrum to characterise
the chemical composition of the atmosphere in order to securely
establish peculiarity or normality. In this section we discuss the
information obtained about each of the stars studied.

\subsection{HD 16605 = NGC 1039-154,  A1p SiSrCr}

The large field of this star was first reported by Kudryavtsev et
al. (\cite{Kudetal06}). Our two measurements confirm their result. All
five available measurements show negative fields, ranging from $-840$ to
$-2430$~G. The observed \vsi\ = 18 \kms\ (see Table~\ref{stars-obs.tab})
suggests a rotation period of a few days. The star is certainly a
magnetic Ap star.

Proper motion data (see Paper II) for this star indicate probable
membership in NGC~1039, as does the star's position near the cluster
isochrone in the HR Diagram. The radial velocity measured for this
star is about -1~\kms.  Our two observations show no variation of this
quantity, and the three spectra obtained by Kudryavtsev et al.
(\cite{Kudetal06}) yield \vr\ values that are consistent with our
measurements (Kudryavtsev, private communication). Thus it appears
that this star is probably not a spectroscopic binary. However, the
radial velocity of the cluster is reported by Kharchenko et al.
(\cite{Khaetal05}) to be $-15 \pm 2.3$ \kms. Furthermore, nearly 50
radial velocities have been measured in low-mass main sequence stars
in this cluster by Jones et al. (\cite{Jonetal97}); omitting two stars
with strongly discordant radial velocities, the median cluster
velocity is about $-10$~\kms with a dispersion of approximately
4~\kms. It is not clear what the origin of the discrepancy between
these two cluster radial velocities is, but since the Jones et al.
data are published, we adopt the median of their values as the cluster
mean \vr. Thus the radial velocity of HD~16605 is about $2 \sigma$
from the cluster median.  This discordant radial velocity is possibly
an indication of non-membership in NGC~1039, but with all other
indicators of membership in favour of membership, we still regard
HD~16605 as a probable cluster member.

\subsection{HD 16728 = NGC 1039-307,  B9V}

No previous magnetic observations of this star are available in the
literature.  We have two observations of the star. One has very low
SNR for a reason that we have not been able to determine, and is
omitted from Table~\ref{fld-meas.tab}. The second
measurement yields a \bz\ value with an uncertainty of almost 300~G,
mainly because the LSD spectral line is shallow and broad, with a
\vsi\ of about 75 \kms. The line profile shape could be that of an
SB2. Synthesis of a small section of the \esp\ $I$ spectrum indicates
that the atmospheric abundances in this star are close to solar
values. (Systematic synthesis of \esp\ $I$ spectra will be a separate
aspect of the investigation of these data which will be reported in a
later paper. Here we simply provide a few first results that help us
to separate magnetic Ap stars from other classes of stars.)  It
appears that this star is probably a normal star of $\te \approx
11500$~K rather than a magnetic Ap star.

Proper motion data indicate that this star is probably a member of
NGC~1039 (Baumgardt et al. \cite{Bauetal00}; Dias et
al. \cite{Diaetal01}). Because it is not clear from our observations
whether the star is single or is a spectroscopic binary, we cannot
test cluster membership with a \vr\ measurement. We consider HD~16728
to be a probable cluster member.

\subsection{HD 19805 = Melotte 20-167,  B9.5V}

A large magnetic field of order 1~kG was reported by Bychkov et
al. (\cite{Bycetal03}), based on four unpublished
measurements. However, three unpublished observations of the star
using MuSiCoS, with \sigbs\ in the range of 125 -- 175~G, do not
reveal any sign of such a large field. Our two observations, with
uncertainties for \bz\ of 24 and 14~G, also show no indication
of any field, nor is a significant Zeeman effect detected in $V$. In
addition, the spectrum of this star closely resembles that of a normal
star of $\te \approx 10200$~K, with strong He~I and O~I, normal Cr~II,
and no obvious sign of rare earths. Synthesis of $I$ in small spectral
regions confirms essentially solar abundances of these elements. We
classify this star as normal.

On the basis of proper motions, this star is a member of the $\alpha$
Per cluster, Melotte 20 (de Zeeuw et al. \cite{deZetal99}). Robichon
et al. (\cite{Robetal99}) give the cluster velocity as $-0.2 \pm
0.5$~\kms; the values tabulated by Kharchenko et
al. (\cite{Khaetal05}) are similar. Our observed radial velocity $\vr
= 0$~\kms is consistent with membership;  we consider this star to
  be a cluster member.

\subsection{HD 108945 = Melotte 111-160,  A2p SrCr}

A weak field ($109 \pm 44$~G) was definitely detected in this star by
Shorlin et al. (\cite{Shoetal02}). One measurement using FORS1 was
reported in Kochukhov \& Bagnulo (\cite{KocBag06}), where the field
value is based only on the Balmer lines; no field was detected. In
Table~\ref{esp-fors-comp.tab} we give the field value measured from
these same ESO archival spectra using both Balmer and metal lines;
again no field is detected, with $\sigb = 75$ G. In two of our three
\esp\ measurements, \bz\ is non-zero at more than the $3 \sigb$
level, although the field is always of order 100 G or
less. Furthermore, in all three measurements, a clearly non-zero
signal is found in $V$, confirming the reality of the small fields
measured. HD~108945 is certainly a magnetic Ap star.

Proper motion data (cf Paper II) indicate that this star is a member
of the Coma Berenices cluster. The mean cluster \vr\ is $-0.1 \pm
0.2$~\kms\ (Robichon et al. \cite{Robetal99}). Our measured radial
velocity of +1 \kms\ is consistent with membership; we consider this
star a cluster member.

\subsection{HD 144661 = HIP 79031,  B8p He-wk}

Three \bz\ observations of this He-weak star, with $\sigb
\sim 200$~G, were made by Borra et al. (\cite{Boretal83}), but no
strong evidence of a field was found. Our two magnetic measurements
($\sigb \sim 50$~G)
yield \bz\ values not significantly different from zero, and no sign
of a magnetic signature is found in either $V$ line profile. The LSD
line profile is constant and appears to be a simple rotational
profile. Borra et al. reported that this star is a member of the PGa
class of He-weak stars, which seems to be a high-temperature extension
of the HgMn stars and, like the HgMn stars, seem to be chemically
peculiar but non-magnetic. Woolf \& Lambert (\cite{WooLam99}) suggest
that the star is Hg-rich. Synthesis of a small section of spectrum
shows that the atmosphere is depleted in He (by more than 1 dex) and
S, almost normal in Si, and about 0.5 dex overabundant in Fe (see also
Norris \cite{Nor71}). We classify HD~144661 as a non-magnetic, He-wk
(probably PGa; cf Borra et al. \cite{Boretal83}) star. 

On the basis of proper motions (de Zeeuw et al. \cite{deZetal99}) the
star is a member of Sco OB2. de Zeeuw et al. give a mean radial
velocity of $-4.6$~\kms\ for the Upper Sco region of the
association. Our \vr\ values are consistent with this value, and we
consider the star an association member.

\subsection{HD 153948 = NGC 6281-15,  Ap Si}

A field of about 200 G was detected in this star using FORS1 (Paper
I). All three of our measurements yield fields of about this size,
although none of the \bz\ values differs from zero by $3
\sigb$. However, two of our observations yield definite detection of
non-zero $V$ signatures in the LSD lines. Furthermore, the line
profile is strongly variable in shape, but not in \vr. We classify
HD~153948 as a definite magnetic Ap star. 

The proper motions are consistent with membership in NGC~6281 (Paper
II). The position in the HR Diagram is consistent with the cluster
isochrone. On the other hand, the radial velocity measured on all
three nights, about $-2.5$~\kms, is inconsistent with the reported
cluster radial velocity of $- 15.4$ \kms\ (Kharchenko et
al. \cite{Khaetal05}). However, this value for the cluster radial
velocity appears to be based on only one measurement of one star, and
is therefore highly uncertain. We disregard this inconsistency, and
accept HD~153948 as a probable cluster member.

\subsection{HD 317857 = NGC 6383-3,  A1IVp}

Both FORS1 (Paper I) and our single field measurement of this star
reveal a strong (kG) field. HD~317857 is certainly a magnetic Ap
star. However, both the radial velocity of $-24$~\kms\ measured (once)
by Lloyd Evans (\cite{LloEva78}), and the value of $-10$~\kms\ from our
\esp\ spectrum, are inconsistent with the cluster radial velocity of
about $+5$~\kms reported by Kharchenko et al. (\cite{Khaetal05});
although the two stellar measurements taken together suggest that the star
is a spectroscopic binary. More definitively, HD~317857 is about 0.8
dex too bright for the cluster isochrone. It appears  that
this star is a probable cluster non-member.

\subsection{HD 318100 = NGC 6405-19,  B9p}

Both the FORS1 measurement and the present \esp\ observation reveal
longitudinal fields of a few hundred G, significant at the several
\sigb\ level. The star also has an extremely large value of the $\Delta
a$ peculiarity parameter (Maitzen \& Schneider
\cite{MaiSch84}). HD~318100 is definitely a magnetic Ap.

Kharchenko et al. (\cite{Khaetal05}) report proper motions that are
consistent with the cluster mean values; our radial velocity of $-8$
\kms\ is consistent with the cluster mean of $-7 \pm 1$~\kms
(Kharchenko et al. \cite{Khaetal05}); and the inferred luminosity
(Paper II) is consistent with the cluster isochrone. We consider
HD~318100 a member of NGC 6405.

\subsection{HD 320764 = NGC 6475-23,  A1V an}

One FORS1 measurement, based mainly on Balmer lines, yielded no field
detection with $\sigb = 66$~G. Our two measurements have uncertainties
of the order of 4--500~G due to the extremely broad and shallow LSD
line, and provide no really useful constraint. Folsom et
al. (\cite{Foletal07}) report \vsi\ = 225 \kms, and their abundance
analysis shows that this is a normal star, not a magnetic Ap. 

Proper motions suggest that this star is  a probable cluster member
(cf Paper II). 

\subsection{HD 162576 = NGC 6475-55,  B9.5p SiCr}

Our three field measurements with \esp\ are the first made of this
star. In spite of standard errors as low as about 20 G, none of the
measurements of \bz\ are significantly different from zero. However,
we have one significant detection of a non-zero $V$ signature. In
addition, the LSD profile is strongly variable, without any change in
\vr. The star has been reported to be photometrically variable (North
\cite{Nor84}), and Folsom et al. (\cite{Foletal07}) report that the
abundances found in the atmosphere are clearly typical of magnetic Ap
stars. We classify HD~162576 as a magnetic Ap star; it appears to be
close to the weak-field limit of such stars (cf Auri\`ere et
al. \cite{Auretal07}). 

Astrometric data indicate that the star is a member of NGC~6475 (cf
Paper II), and our measured values of \vr\ are consistent with
the cluster radial velocity of $-14.7 \pm 0.2$~\kms\ (Robichon et
al. \cite{Robetal99}). This star is  a probable cluster member. 

\subsection{HD 162588 = NGC 6475-59,  Ap Cr}

None of our three field measurements (the first such observations of
this star) reveal a \bz\ value that is significantly different from
zero, but the \sigbs\ are fairly large, of order 100~G, because of
the elevated value of \vsi\ (70~\kms). No significant $V$ signal is
detected. However, the LSD $I$ line profile is strongly variable
without significant change in \vr. The star is a photometric variable
(North \cite{Nor84}). We have synthesised a small section of the $I$
spectrum, and find that Cr is overabundant by about 1.5~dex.  In spite
of the absence of direct evidence for the presence of a magnetic
field, we classify the star as a magnetic Ap star.

Astrometric data indicate that HD~162588 is a member of NGC~6475 (Dias
et al. \cite{Diaetal01}). Our measured \vr\ values are consistent with
the cluster velocity, and  we consider the star a cluster member.

\subsection{HD 162630 = NGC 6475-63,  B9V}

No significant \bz\ field was detected in two measurements, and no
significant $V$ signature is present. The LSD profile is considerably
deeper using a solar line list than with an Ap line list of the same
temperature, which is usually a symptom that many lines present in the
Ap list but not in the solar list are actually weak or absent in the
analysed star. In addition, the LSD line profile is apparently a
simple rotation profile, and it does not vary between our two
observations.  In the $I$ spectrum, He~I and O~I are quite strong,
which is normally not the case for magnetic Ap stars. Synthesis of a
small section of $I$ spectrum shows that Fe appears underabundant
relative to solar by about 0.3~dex, while Si and He have abundances
close to solar. We conclude that HD~162630 is not a magnetic Ap star,
but an (approximately?) normal star of $\te \approx 10400$~K.

HD~162630 does appear, both on the basis of proper motions and of our \vr\
measurements, to be a member of NGC~6475. 

\subsection{HD 162656 = NGC 6475-72,  B9V}

Our single measurement, the first field measurement of HD~162656,
shows no sign of a magnetic field either in the value of \bz\ or in
the LSD $V$ profile. The LSD profile is deeper, and the field
measurement more accurate, using a solar line list than with an Ap
list. The LSD line profile appears to be broadened simply by
rotation. Although the star has a \te\ value of about 9500~K, the $I$
spectrum shows strong He~I and O~I lines, and the Cr~II lines near
6150 \AA\ are weak or absent. Synthesis of short sections of the $I$
spectrum confirms these qualitative observations; He, Si, and Fe all
have abundances close to solar, while Cr appears to be about 0.2~dex
below solar. We classify this star as normal.

The proper motions are consistent with membership in NGC~6475, but our
one measured \vr\ value is not. However, it has been reported by
Gieseking (\cite{Gie77}) that the star is a spectroscopic binary, and
that the $\gamma$ value of the orbit is consistent with cluster
membership.  We consider the star to be a cluster member.

\subsection{HD 162725 = NGC 6475-88,  A0p SiCr}

Our data are the first magnetic field measurements of this star. One
of our two measurements reveals a \bz\ value $7 \sigb$ different
from zero, although the other is consistent with zero. Both
observations have LSD $V$ signatures that are clearly different from
zero. Folsom et al. (\cite{Foletal07}) report that the atmospheric
composition of HD~162725 is clearly that of an Ap star. The star is
also a photometric variable (North \cite{Nor84}). It is clear that
this star is a magnetic Ap star. 

Both proper motions and our radial velocity are consistent with
classification of the star as as member of NGC~6475. 

\subsection{HD 169842 = NGC 6633-39,  A1p SrCr}

Kudryavtsev et al. (\cite{Kudetal06}) report probable detection of a
field of $B_{\rm rms} \approx 370$~G from five measurements with
typical standard errors of about 180~G. Our three observations provide
clear detections of \bz\ at about the $10 \sigb$ level, and confirm
the general magnitude of the field reported by Kudryavtsev et
al. Synthesis of two spectral windows (cf. Figure~\ref{hd169842-458})
indicates that Cr is overabundant relative to solar abundances by
about 2~dex, while Si and Fe are near solar abundances. HD~169842 is
definitely a magnetic Ap star.

Proper motions are consistent with membership in NGC~6633. The cluster
radial velocity is reported by Kharchenko et al. (\cite{Khaetal05}) to
be between $-25$ and $-29$~\kms. Individual radial velocities from
references cited in WEBDA support a value around $-29$~\kms. Our
measured \vr\ values are consistent with these values, and  we
  consider the star to be a cluster member.

\subsection{HD 169959A = NGC 6633-58,  A0p Si}

The field of this star was first detected with FORS1 (Paper I). Our
field measurement confirms the presence of a field and its general
magnitude in this magnetic Ap star. 

The proper motions and previous radial velocity measurements (Hintzen
et al. \cite{Hinetal74}) are consistent with membership in
NGC~6633. However, our one radial velocity (-53 \kms) is inconsistent with
both the cluster velocity of about $-30$~km/s and with the measurements
of Hintzen et al. It is not at all clear what the origin of this
discrepancy is. Furthermore, HD~169959A, if it is a cluster member, is
a (very) blue straggler. We conclude that it is  a probable
cluster non-member. 

\subsection{HD 170054 = NGC 6633-77,  B6IV}

No field was detected in two previous FORS1 measurements with standard
errors of about 75~G (Paper I). Our two new measurements, one with a
larger \sigb\ (due to a period of low transparency) and one with
smaller \sigb\ also reveal no field, nor is there any hint of a field
in the $V$ signatures. Spectral lines of He~I, O~I, and Ne~I are
strong, while Cr and Fe appear roughly normal. Synthesis confirms
abundances of He and Fe that are close to solar, while Si appears
about 0.2~dex below solar. We classify HD~170054
as a normal B star, although the peculiar shapes of the $I$ line
profiles, and their variability, suggest that this star may be a
pulsating variable. 

Proper motions indicate that the star is a cluster member, and our
\vr\ value is consistent with this.  We consider the star to be a
  cluster member.

\subsection{HD 170860 = IC 4725-153,  B9IV/Vp}

The magnetic field of this star was not detected in one FORS1
measurement, which achieved $\sigb = 69$~G. Our single \esp\
measurement nominally provides about a $2 \sigb$ detection, although
both \bz\ and \sigb\ are not very well determined, due to
uncertainties as to where the edges of the LSD line profile
are. However, a large and clearly significant $V$ signature is
present. In addition, the $\Delta a$ value (Maitzen \cite{Mai85}) is
rather large, providing further support to our classification of this
star as a magnetic Ap star.

Proper motions indicate that HD~170860 is a member of IC~4725. 
  The mean cluster radial velocity has been reported as having values
  from $-4 \pm 4$ (Feast \cite{Fea57}) to +2.4~\kms (Kharchenko et
  al. \cite{Khaetal05}). Because of the difficulty of determining the
centroid of the LSD line, our \vr\ value is not derived from this
line; instead, it is derived by fitting Gaussians to the cores of the
Balmer lines H$\alpha$ -- H$\delta$, and the strong Si~II lines
4128-30 and 6347-71 \AA. The value found by us, $-13 \pm 3$~\kms, is
not the same as  the cluster mean, but with only one measurement,
we cannot rule out the possibility that the star is a spectroscopic
binary. We regard HD~170860 as a probable cluster member.

\subsection{HD 172271 = IC 4756-118,  A0p Cr}

Our measurement is the first reported magnetic observation of this star. We
find a marginally significant \bz\ in this star. The significantly
peculiar value of the Geneva index $Z = -0.027$, and the apparent
absence of He~I lines in HD~172271, strongly suggest classification
as a magnetic Ap star. This is confirmed by synthesis of a small
spectral region, from which we find that Cr is overabundant by about
2.5~dex, while Fe is about 0.5~dex overabundant. 

The value of $\mu_\alpha \cos \delta$ is about $2\sigma$ from the
cluster mean, but $\mu_\delta$ is consistent with membership. The
cluster \vr\ is $-25.8 \pm 0.2$~\kms\ (Robichon et
al. \cite{Robetal99}; note sign error on their \vr\ value); our
measured $\vr = -21 \pm 2-3$~\kms (the star has a large \vsi) is
reasonably consistent with this value. We regard this star as a
probable cluster member.

\subsection{HD 205073 = NGC 7092-69,  A1}

Our two measurements of this star are the first reported magnetic
observations. No field is detected, either in the \bz\ values or in
the $V$ profiles. Considering the rather small \vsi\ value (16~\kms), the
lack of detection is strong evidence that this star is not a
magnetic Ap. Furthermore, the field measurement is more precise using
a solar composition line list for LSD analysis. Neither the $\Delta a$
nor the Geneva $Z$ peculiarity indices are significantly different
from zero, and our spectra show line profiles which appear to be
simply rotationally broadened, with clearly present He~I and O~I but
very weak Cr~II and no rare earths. The abundances of Cr and Fe
deduced from spectrum synthesis of small sections of the spectrum are
close to solar. It is clear that this star is a normal A star. Note
incidentally the two quite different \vr\ values of the two
observations; HD~205073 is a spectroscopic binary. Since close
binaries are rare among magnetic Ap stars, the radial velocity
variability supports our view that HD~205073 is not such a star.

The proper motions are consistent with cluster membership (Baumgardt
et al. \cite{Bauetal00}; Dias et al. \cite{Diaetal01}).  We consider
the star to be a cluster member.

\subsection{HD 205331 = NGC 7092-118,  A1V}

The first published magnetic measurements of this star were those of
Kudryavtsev et al. (\cite{Kudetal06}), who failed to detect a field in
three observations with \sigb\ of about 280~G. Our three data reduce
the standard error to the $30 - 40$~G range. We do not have a
statistically significant detection in the $V$ signature in any of the
three observations, but one \bz\ value (integrated over the whole
broad line) is about $3\sigb$ different from zero, and suggests a
field of about 140~G. However, our field measurement is as precise
with a solar line list as with an Ap list, and the LSD profile appears
to be due to pure rotational broadening. Both these features are often
signatures of a normal star. Our spectra show a clear He~I $\lambda
5875$ line (in spite of the low \te\ value), strong O~I, weak Cr~II,
and no obvious rare earths. The brightness is constant in the
Hipparcos photometry catalogue. Synthesis of two short spectral
regions in the \esp\ $I$ spectrum yields near-solar abundances of Si,
Cr and Fe, assuming that the microturbulence parameter is about
2~\kms. We classify HD~205331 as a normal star of $\te \approx 9450$~K
in spite of the single apparently significant field detection.

The proper motions and parallax are consistent with cluster membership
(Baumgardt et al. \cite{Bauetal00}; Dias et al. \cite{Diaetal01}).
The accurate cluster radial velocity of $-5.4 \pm 0.4$~\kms\ (Robichon
et al. \cite{Robetal99}) is agrees with the three values of \vr\
measured by us. We consider the star a cluster member.

\subsection{BD +49 3789 = NGC 7243-490,  B7p Si}

Our single measurement of the field of this star is the only one
available. The precision is very low ($\sigb \sim 400$~G), and no
detection is achieved. The large values of the photometric peculiarity
indicators $\Delta a$ and Geneva $Z$, and the presence of strong Si~II
lines in our $I$ spectrum suggest that BD +49 3796 is actually a
magnetic Ap star. This view is confirmed by synthesis of two small
spectral windows, which yields an underabundance of He by about 1~dex,
and overabundances of Si, Cr and Fe by respectively 1, 0.7, and
0.5~dex relative to solar abundances.

Astrometric data for this star, and HR Diagram position, are
consistent with cluster membership, but our single \vr\ value differs
from the cluster mean of between $-9$~\kms\ (Hill \& Barnes
\cite{HilBar71}) and $-13$~\kms\ (Kharchenko et al. \cite{Khaetal05})
by about 5 \kms. However, Hill \& Barnes report four \vr\ measurements
between $-13$ and $+32$~\kms, and consider the star a spectroscopic
binary. We regard the star as a probable member.

\subsection{HIP 109911 = NGC 7243-370,  A0Vp Si}

Our magnetic measurement is the only one available for this star, and
does not reveal a significant \bz\ or $V$ signal, although there is a
hint in our low-precision $2\sigb$ \bz\ that there might be a field
of a few hundred~G present. In any case, the photometric peculiarity
indices $\Delta a$ and Geneva $Z$ are large enough to strongly suggest
that HIP~109911 is a magnetic Ap, and the obvious strength of Si~II
and Cr~II lines supports this view. Synthesis of a small spectral
window yields He depleted by 1.5~dex, and Si and Fe enhanced by 1 and
1.2~dex respectively relative to solar. We classify the star as a
magnetic Ap star.

The proper motions (Baumgardt et al. \cite{Bauetal00}; Dias et
al. \cite{Diaetal01}) and position in the HR Diagram
are consistent with cluster membership. Hill \& Barnes
(\cite{HilBar71}) report four \vr\ values between $-8$ and $-24$~\kms\
but do not flag the star as a velocity variable, perhaps reflecting
the relatively large mean errors. Our single \vr\ value, with an
uncertainty of roughly 2~\kms, is consistent with membership.  We
  consider the star a cluster member.

\section{Characteristics of the data set}

In three nights (of which all but a few hours were clear) with \esp\
we have obtained 44 observations of 23 stars which are possible
magnetic Ap stars {\em and} possible members of open clusters. For 12
of these 23 stars, our measurements represent the first published
magnetic data.

Because of the time of year when our observing run was scheduled,
about half of our \esp\ observations were made on stars which had been
previously observed, either earlier in our cluster survey with FORS1
(Paper I), or by others. Nevertheless, the data we have acquired
provide valuable new information about possible cluster magnetic
stars, even when these stars have been previously measured for
magnetic fields.

One important class of information provided by our new \esp\ data is
accurate radial velocities of almost all the stars observed. Because
the usefulness of proper motions as a discriminant of cluster
membership diminishes roughly proportionally to distance (the required
accuracy of motions increases with distance, but the uncertainties of
available data set remain roughly constant), while radial velocities
continue to be a precise discriminant to whatever distance they can be
measured accurately, the radial velocities are particularly useful in
testing membership of stars in clusters that are more than 1 -- 200 pc
distant.  For all but a few of the 23 stars observed, our data provide
a valuable test of membership, mostly supporting cluster
membership. However, our \vr\ data indicate that two stars previously
thought to be cluster members  (HD~317857 and HD~169959A) are
probable non-members.


Another important type of information provided by \esp\ is clearer
discrimination between stars which are magnetic Ap stars with fields
small enough to be difficult to detect, and stars which are not
magnetic Ap stars. In order to obtain representative field strength
{\em distributions} for various star masses and ages, we have included
a number of stars in which fields have not yet been detected, but for
which other evidence exists indicating that they are probably magnetic
Ap stars, in the sample analysed in Paper II. Even with \esp\
observations of such stars it is not always possible to detect or rule
out a field, but in most cases the \esp\ spectrum allows one to
classify a star with confidence as either a magnetic Ap or a normal
star, something which is not usually possible with FORS1 $I$ spectra
(see for example the discussions of HD~162630, HD~162656, HD~170054,
HD~205073, and HD~205331). Our $I$ and $V$ spectra allow us to
categorise nine stars of the present sample as not magnetic Ap stars;
in most cases, these appear to be normal stars.


One obvious aspect of the present data set is that in stars with
relatively rich and sharp-lined metal spectra, \esp\ data may allow
the detection of fields not readily detected with FORS1. Of the two
stars for which we report the first direct evidence of a magnetic
field, one (HD~162725) was previously observed with FORS1 (Paper I).

Of course, for all the stars we observe, the new measurements
contribute to convergence of \brms, the RMS longitudinal field (see
Paper II and below), which is the measure we use of the global
characteristic field strength of each star in our sample.

\section{Magnetic field evolution through the main sequence}

\begin{table*}
\begin{center}
\caption{\label{cl_stell_params_magmeas.tab} Physical properties of stars that are probable or certain open cluster members, probable or certain Ap stars, and have new ESPaDOnS magnetic field measurements.}
\begin{tabular}[c]{lrrlrrrrrrrrr} 
\hline 
\hline \\ 
Cluster  &   $\log t$ &  true DM &    Star      & $\log T_e$ & $\log L/L_\odot$ &       $M/M_\odot$   & fractional age       & $B_{\rm rms}$  & $n_B$ \\    &  (yr)  &   &   &  (K)  &   &   &   &  (G) \\ 
\hline \\ 
NGC 1039      &   8.33 $\pm$   0.06 &   8.49  &   HD 16605      &  4.025  &         1.65      &   2.55 $\pm$   0.15 &    0.40 $\pm$   0.09 &    1894 &    5 \\ 
Coma Ber      &   8.70 $\pm$   0.10 &   4.70  &   HD 108945     &  3.944  &         1.60      &   2.30 $\pm$   0.15 &    0.71 $\pm$   0.21 &      93 &    4 \\ 
NGC 6281      &   8.45 $\pm$   0.10 &   8.86  &   HD 153948     &  4.025  &         1.86      &   2.70 $\pm$   0.15 &    0.62 $\pm$   0.18 &     201 &    4 \\ 
NGC 6405      &   7.80 $\pm$   0.15 &   8.44  &   HD 318100     &  4.025  &         1.64      &   2.50 $\pm$   0.10 &    0.11 $\pm$   0.04 &     517 &    2 \\ 
NGC 6475      &   8.41 $\pm$   0.05 &   7.24  &   HD 162576     &  4.004  &         2.17      &   3.10 $\pm$   0.15 &    0.81 $\pm$   0.17 &      15 &    3 \\ 
              &                 &         &   HD 162588     &  4.033  &         2.10      &   3.05 $\pm$   0.20 &    0.77 $\pm$   0.17 &      34 &    3 \\ 
              &                 &         &   HD 162725     &  3.982  &         2.36      &   3.30 $\pm$   0.20 &    0.94 $\pm$   0.20 &      69 &    3 \\ 
NGC 6633      &   8.75 $\pm$   0.05 &   7.93  &   HD 169842     &  3.924  &         1.62      &   2.35 $\pm$   0.10 &    0.85 $\pm$   0.18 &     315 &    8 \\ 
IC 4725       &   8.02 $\pm$   0.05 &  10.44  &   HD 170860     &  4.137  &         2.63      &   4.25 $\pm$   0.20 &    0.71 $\pm$   0.15 &      40 &    1 \\ 
IC 4756       &   8.44 $\pm$   0.06 &   7.59  &   HD 172271     &  4.013  &         1.98      &   2.85 $\pm$   0.15 &    0.69 $\pm$   0.16 &     265 &    1 \\ 
NGC 7243      &   8.06 $\pm$   0.10 &   9.54  &   BD+49 3789    &  4.111  &         2.23      &   3.55 $\pm$   0.15 &    0.51 $\pm$   0.15 &     561 &    1 \\ 
              &                 &         &   HIP 109911    &  4.114  &         2.29      &   3.65 $\pm$   0.15 &    0.54 $\pm$   0.16 &     494 &    1 \\ 
\hline 
\hline 
\end{tabular} 

\end{center}
\end{table*}

We next use the new data acquired from \esp\ to revise our global
database of magnetic Ap stars that are members of clusters. We flag
stars that are found to be probable or certain non-members, or
probable or certain non-magnetic stars. We add newly studied probable
or certain magnetic cluster Ap stars to the database. For each of the
new stars, we determine the fundamental parameters \te\ and \logl\
using the same techniques as in Paper II. As discussed in detail in
that paper, we assume that the uncertainty in \logte\ is $\pm 0.02$,
and the uncertainty in \logl\ is $\pm 0.1$. With these data, we place
each star in the cluster HR diagram, check whether its position in the
HR diagram is close enough to the cluster isochrone to be consistent
with membership, and estimate its mass from its position.

For each retained magnetic cluster member (old or new) in the database
for which we have new magnetic data, we compute the revised
characteristic field strength \brms, which is simply the RMS average
of all available field measurements, or the subset with the smallest
uncertainties. (Although \brms\ is a fairly crude measure of stellar
field strength, it is the best measure we can easily form from the
data until enough is known about most stars in the sample to actually
compute simple models of individual field structure.)  Finally, we
re-examine the evolution of magnetic fields with time elapsed on the
main sequence to see if any of the conclusions of Paper~II have to be
modified.

After carrying out these steps, we find that we have 12 stars which
satisfy our criteria of cluster membership and probable or definite
magnetic Ap nature, and for which the magnetic field data are new or
improved, so that a new value of \brms\ may be calculated. These stars
are listed in Table~\ref{cl_stell_params_magmeas.tab}. The table is
similar to Table~3 of Paper II in the information it contains. For
each magnetic Ap cluster member for which new magnetic data have been
obtained, the table lists the cluster to which the star belongs, the
cluster age and uncertainty (see below), the true (not apparent)
distance modulus $DM$ of the cluster, the star identification, the
star's \logte, \logl, \mmo\ with uncertainty, fractional age $\tau$
with uncertainty, the revised value of \brms, and the number $n_B$ of
magnetic observations contributing to \brms. (Note that for Ap stars
for which we have not yet detected a field, we use only the most
accurate measurements in forming \brms, so as not to inflate that
parameter by including numbers with unnecessarily large uncertainties.)

We also use the newly measured stars to test the cluster ages found in
the literature by comparing the position of stars near the terminal
age main sequence (TAMS) with the isochrones. In Paper II, we found
that this procedure made it possible to improve the precision of age
determination of several open clusters. Compared to Paper II, we
obtain improved ages with smaller error bars for NGC~6475,
IC~4725, and IC~4756. These improved ages are listed in column~2 of 
Table~\ref{cl_stell_params_magmeas.tab}.   

\begin{figure*}
\resizebox{16.0cm}{!}{    
\includegraphics*[angle=0]{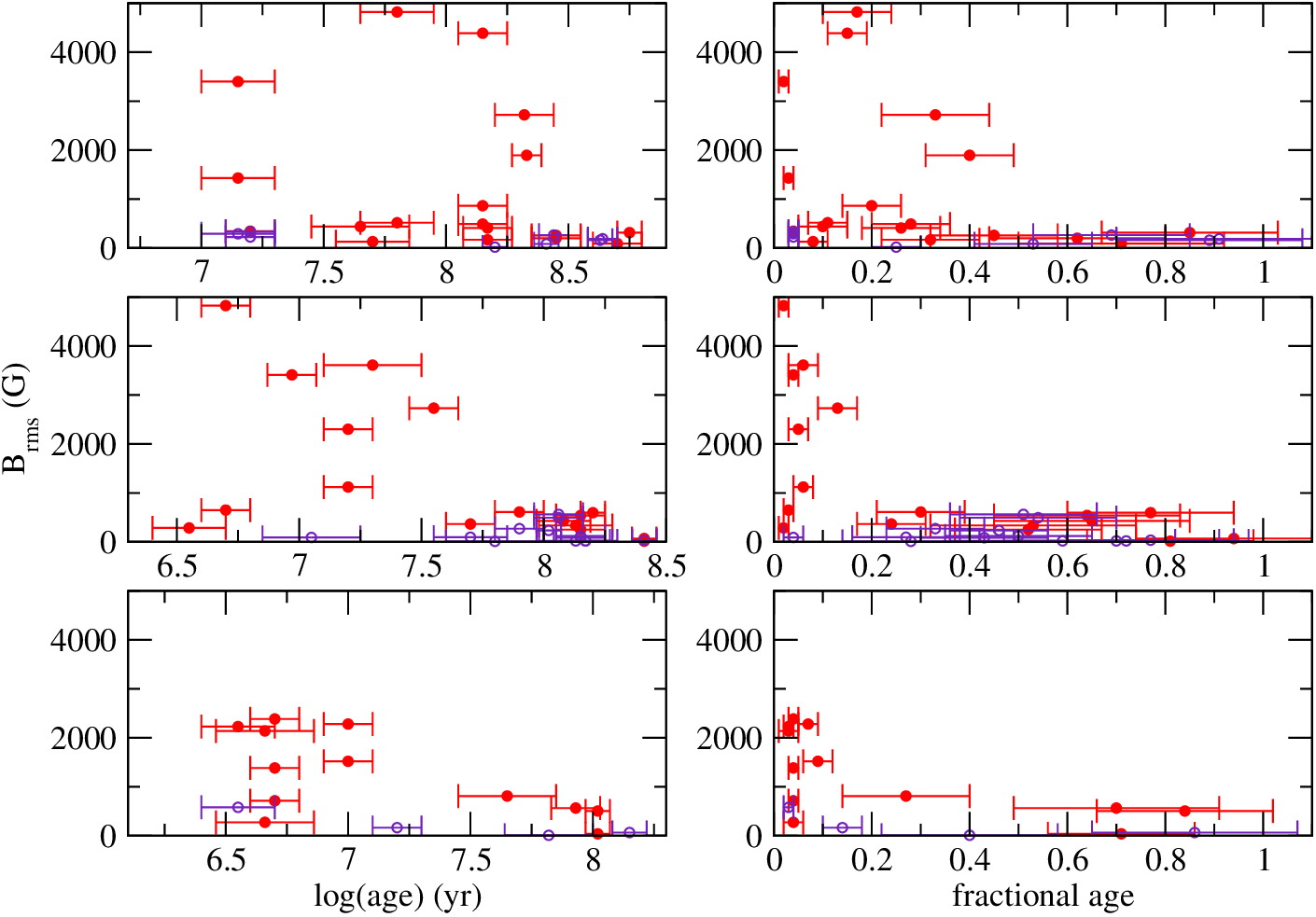}
}
\caption{\label{field-vs-age.fig} This figure shows the currently
  estimated values of \brms\ as functions of logarithmic stellar age
  (left) and of fractional age (right) for three mass bins, from top
  to bottom $2 - 3 \mo$, $3 - 4 \mo$, and $4 - 5 \mo$. Filled symbols
  are stars for which a field is definitely detected; open symbols are
  probable magnetic Ap stars in which no field has yet been
  detected. The right-hand limit of each of the panels using log(age)
  as abscissa is near the main sequence lifetime for stars in that
  mass range. In the bottom pair of panels, one point (for NGC
  2244-334) has such a large field ($\brms = 9.52$~kG) that it is off
  scale (at $\log t = 6.4 \pm 0.10$ and $\tau = 0.02 \pm 0.01$
  respectively). 
  }
\end{figure*}

Then finally we include the new data together with the data from
Paper~II to check whether the new magnetic measurements alter our
previous conclusions (Paper II) concerning the evolution of magnetic
fields with time through the main sequence lifetime of magnetic Ap
stars. The full, revised data set of stars which are probable cluster
members, probable Ap stars, and for which at least one field
measurement is available, is used to create
Figure~\ref{field-vs-age.fig}, which contains information similar to
that found in Figure~5 of Paper~II.

In this figure the RMS field estimates \brms\ for individual stars (in
three mass bins: 2 -- 3 \mmo, 3 -- 4 \mmo, and 4 -- 5 \mmo, bins for
which a significant number of stars are now available) are plotted as
a function of cluster (and thus of stellar) age (left three panels)
and of fractional age (right three panels). The full sample is divided
into mass bins in order to be able to study the variation of typical
field strength \brms\ with stellar age as a function of stellar
mass. (Note that the star NGC~2244-334, with $\brms = 9515$ G, log(age)
= 6.40, and $\tau = 0.02$ is off-scale in the  upper left corner of
both bottom panels.)

\begin{figure*}
\resizebox{16.0cm}{!}{    
\includegraphics*[angle=0]{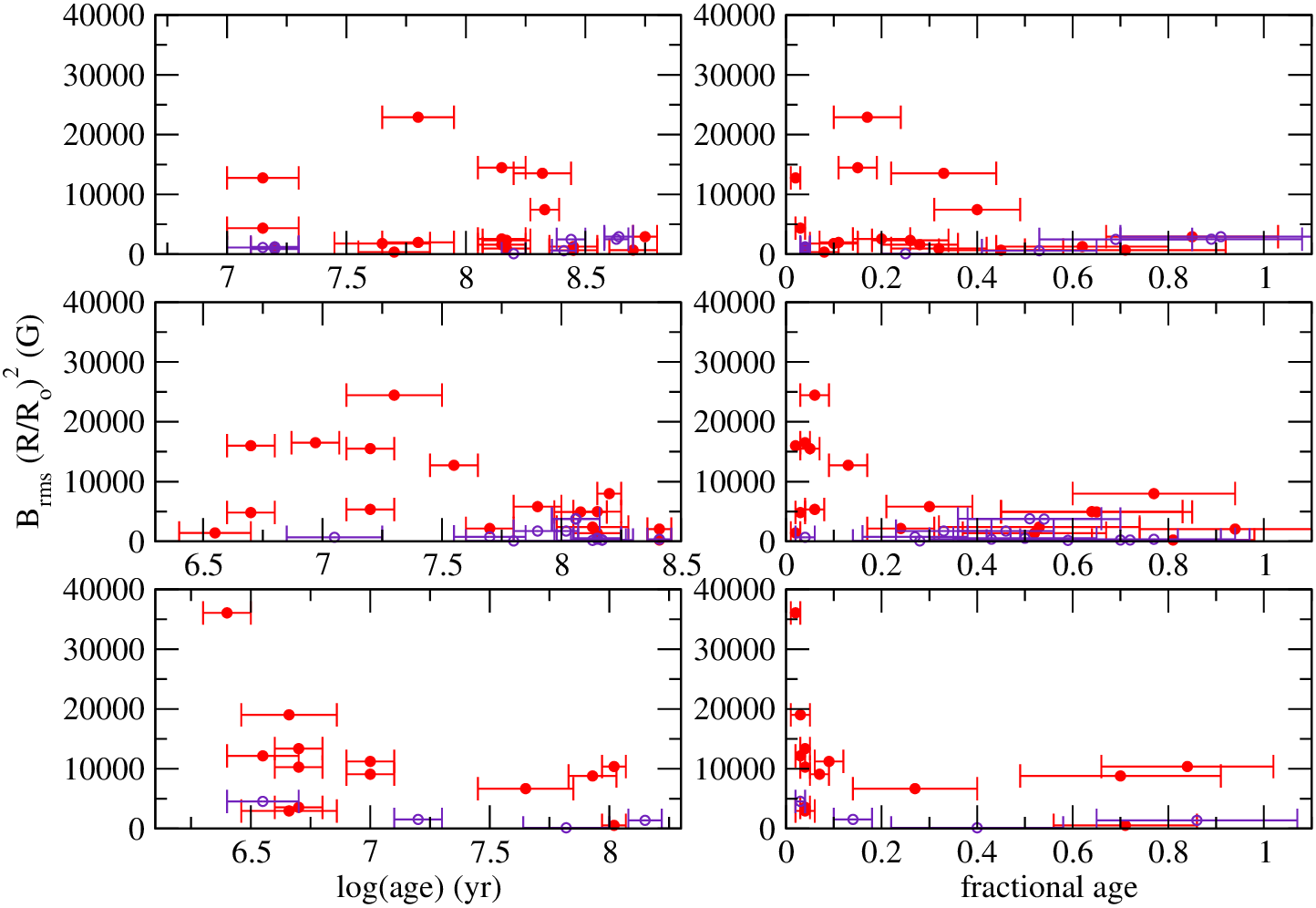}
}
\caption{\label{flux-vs-age.fig} This figure shows the estimated
  values of normalised emergent magnetic flux $\sim \brms \times
  (R/R_\odot)^2$ as functions of logarithmic stellar age (left) and of
  fractional age (right) for three mass bins. Apart from the quantity
  plotted, this figure has the same structure as
  Figure~\ref{field-vs-age.fig}.}
\end{figure*}

The present sample differs from the sample of Paper~II essentially in
that the first magnetic measurements for five stars have been
added, and the \brms\ data for five other stars have changed, always
by a factor of less than two. Although the \esp\ data do not increase
the total size of the sample very much, these observations have
substantially increased the information available about lower-mass Ap
stars that have completed a large part of their main sequence
evolution (see the fractional age column of
Table~\ref{cl_stell_params_magmeas.tab}). These data provide a
significant improvement for a region of parameter space that was not
very well represented in the previous sample.


The data of the full sample of probable cluster magnetic Ap stars are
displayed here a little differently than in Paper~II. In
Figure~\ref{field-vs-age.fig} we show the variation of \brms\ (rather
than its logarithm) as a function of stellar (= cluster) age in the
left panels, and as a function of fractional age in the right
panels. The three panels in each column show the data for the three
mass bins mentioned above. In all three right-hand panels we can now
see clearly that RMS fields larger than 1 kG are common at small
fractional age, while the (now numerous) stars of fractional ages
larger than about 0.4 for 2 -- 3 \mo, 0.2 for 3 -- 4 \mo, and 0.1 for
4 -- 5 \mo\ have fields that are always less than 1 kG. These trends
are not so easily seen in the left panels using log(age) as the
abscissa, which expand the region of small fractional age at the
expense of regions of large fractional age, but close examination of
the left panels shows that the same effect is present. In all three
panels on the left side of the figure, a significant number of low
field stars (but no high-field stars) are present near the TAMS.

The obvious interpretation of this figure is that the RMS field
declines rather sharply with age after a characteristic time which is
of the order of 250 Myr for stars in the mass range 2 -- 3 \mo, 40 Myr
for 3 -- 4 \mo, and 15 Myr for 4 -- 5 \mo stars. It is remarkable that
the characteristic time of the field decreases rapidly as the stellar
mass increases, {\em both} in absolute value and as a fraction of the
stellar main sequence lifetime.

One expected physical effect which would lead to some substantial
decline in surface magnetic field is of course the geometric expansion
of the stars by a factor of about two in radius as they evolve from
the ZAMS to the TAMS. If magnetic flux (and global field topology) are
approximately conserved, the increase of radius by a factor of order
two will lead to decrease in field strength by a factor of order
four. We test this possibility by plotting $\brms \times
(R/R_\odot)^2$ for the stars of Figure~\ref{field-vs-age.fig} as a
function of log(age) and $\tau$ in Figure~\ref{flux-vs-age.fig}. This
quantity is a reasonable proxy for the emergent or apparent magnetic
flux (normalised to an average field of 1~G over a star of one
$R_\odot$); it differs from the true normalised emergent flux by a
factor (dependent on the viewing geometry and obliquity of the stellar
field) which relates the RMS value of the observed sample of values of
mean longitudinal field \bz\ to the typical field modulus at the
stellar surface. This factor is at least about 3, and is likely to be
of the order of 5 for most of the stars of our sample. Of course, the
emergent flux may or may not be closely related to the flux deep in
the stellar interior.  However, note that fields {\em are}
  detected among the older stars in all our mass bins. The fields
  decline in all bins, but not to zero in the main sequence
  lifetime of the stars considered.

Overall we see that Figure~\ref{flux-vs-age.fig} shows much the same
behaviour as Figure~\ref{field-vs-age.fig}. Near $\tau = 0$ there is a
rather large range of estimated normalised emergent magnetic fluxes,
and a few stars which exceed the median normalised flux value by a
factor of 10 or more. As one moves towards larger fractional age, the
fluxes converge on lower values -- quickly for the higher mass ranges,
more slowly for the (top) 2 -- 3 \mo\ bin. From the right column of
figures, one plausible hypothesis is that stars with initially low
fields, less than about 1~kG, tend to conserve flux as they evolve
across the main sequence. However, the stars with initially large
fields, those that contribute the large normalised flux values in the
figure, appear to lose much of this flux within a time short compared
to the main sequence lifetime. This flux loss is {\em not} due to the
geometric expansion of the star, already taken into account in the
method of estimating the normalised magnetic flux. Instead, it could
be some kind of relaxation or redistribution process 

We do not find any analogues of the young, high-field (and high
emergent flux) stars among the (now reasonably numerous) older stars
in any of the mass bins considered here. We conclude that the present
data may be consistent with approximate emergent flux conservation in
most magnetic Ap stars, but that  the initially high-field stars
seem to suffer a real decline in the total emergent magnetic flux
during the main sequence phase of evolution. Alternatively, the
present data may show that  typical emergent magnetic flux
  declines significantly in most or all stars (rather than only in the
  stars with initially large fields and fluxes), by a factor of two or
  three, during the main sequence phase.

 It appears that our observations are consistent with the fossil
  field hypothesis (as we have implicitly assumed in discussing the
  possibility of flux conservation above), in that the flux in older
  stars are within a factor of two or three of the fluxes in young
  stars in each mass bin. As many authors have noted, these fluxes do
  not show the expected correlation with rotation that is found in the
  presumably dynamo-generated fields of lower main sequence stars. In
  fact, some of the largest fields are found among stars that rotate
  very slowly. 

  Furthermore, one theoretical concern about the fossil field
  hypothesis has be the question of whether fossil fields are stable
  over long time scales. Tayler (\cite{Tay73}), Markey \& Tayler
  (\cite{MarTay73}; \cite{MarTay74}), and Wright (\cite{Wri73}) have
  shown that fields which are purely poloidal or purely toroidal in the
  radiative interior of a star are unstable on time-scales that are
  short compared to the main sequence lifetime of an Ap star. The fact
  that we find fields with similar fluxes at all ages in each mass bin
  certainly suggests that, if these fields are fossils, nature has
  been able to resolve this problem.

  This problem has recently been clarified on the basis of a set of
  important calculations discussed by Braithwaite \& Spruit
  (\cite{BraSpr04}) and Braithwaite \& Nordlund
  (\cite{BraNor06}). This study has been primarily focussed on the
  issue of the long-term stability of fossil magnetic fields in Ap
  stars (and magnetic white dwarfs). These authors conclude, on the
  basis of numerical simulations of field evolution in a non-rotating
  polytropic star, that the instabilities found in earlier work lead
  to rapid field evolution which generally creates a mixed
  poloidal-toroidal field structure with a twisted toroidal
  field. Once formed, this structure appears to be stable on a longer
  time-scale. This work appears to provide confirmation that stable
  fossil field structures exist, and can be achieved by a star starting
  from a variety of initial fields, in agreement with the
  observations of fields with comparable fluxes (to within plus or
  minus an order of magnitude) in Ap stars of all main sequence ages.

  Since this work studies the evolution of field structure in
  radiative stars similar to main sequence stars, it also leads to
  predictions of how a global fossil magnetic field should evolve with
  time, at least under the hypotheses adopted in the model. The
  evolution found by Braithwaite and collaborators, when their
  computations are rescaled to approximately realistic magnetic
  diffusivities, is that the emergent flux gradually grows with time,
  by a factor that may be of order ten, until finally the main
  toroidal flux tube begins to emerge from the star, at which point
  the surface field declines rapidly. (The emergent flux is mostly
  poloidal, since a stellar atmosphere does not support the large
  current densities needed for a strong surficial toroidal field
  component.) This evolution is estimated (roughly) to take somewhat
  longer than the main sequence lifetime of most Ap stars. Thus these
  computations predict that we should mainly see stars in the phase of
  surface field strength {\it increasing} with time.

  This is, of course, the opposite of the situation revealed by our
  data, in which we see both field and flux decline with time, at
  least in some stars and perhaps in all. Thus it appears that the
  field evolution predicted by the numerical modelling of Braithwaite
  and collaborators is not fully consistent with observations. The
  observations {\it are} consistent in that fields appear to be stable
  on a long enough time scale to persist throughout the main sequence
  phase, in agreement with the numerical models, but the observed and
  computed fields appear to have somewhat different time
  evolution. 

  However, it is perhaps not very surprising that the field evolution
  predicted by these numerical models differs from what is
  observed. This modelling neglects two effects that may significantly
  affect the evolution of surface fields. First, the computations are
  for {\it non-rotating} stars. The effect of reasonably rapid
  rotation (and most magnetic Aps, including most of the stars in our
  study, have rotational periods of a few days, and equatorial
  velocities of several tens of \kms) is to produce meridional
  circulation. In turn, this circulation transports angular momentum
  and almost certainly produces shear within the star. This shear will
  certainly strongly distort the internal field, which will probably
  substantially alter the emergent flux. Secondly, the modelling has
  been carried out for a star of constant size and structure. During
  the main sequence phase, a real star changes its overall structure
  significantly, and in particular increases in radius by a factor of
  order two. This radius increase will counteract to some extent the
  increases surface field strength (although not the increased surface
  flux). It appears that further computations of field evolution,
  with incorporation of additional physics, will be of great
  interest.

\section{Conclusions}

In order to study the evolution of magnetic field strength and of
atmospheric chemistry in magnetic Ap stars through their main
sequence lifetimes, we have been carrying out a large survey of such
stars in open clusters. As discussed in Papers~I and II, the main
interest of such a sample is the fact that the ages of such stars,
especially during the first half of their main sequence lifetime, can
be determined with considerably greater precision than is the case for
magnetic Ap stars in the nearby field. 

Most of the survey so far has been carried out using FORS1 at the ESO
VLT (cf Paper~I), with an important contribution from earlier studies
of the Ori OB1 and Sco OB2 associations (see for example Borra
\cite{Bor81} and Thompson et al. \cite{Thoetal87}). In this paper we
discuss the first use of the new instrument \esp\ at the
Canada-France-Hawaii Telescope for this survey. The three nights of
data so far obtained from this instrument have made an important
additional contribution to the overall survey. They have also made
possible several very important checks of previous data, and provided
kinds of information not readily available from the lower resolution
FORS1 spectra.

Comparison of \esp\ and FORS1 measurements of the same stars show that
for stars that are not too hot (say below about 12--13\,000~K) and
that have \vsi\ less than roughly 60 \kms, the precision in \bz\
obtainable with \esp\ in a given time on target is comparable to, and
sometimes substantially better than, the precision obtained with
FORS1, even though the CFHT has only about 20\% of the collection area
of one VLT unit telescope. This is possible essentially because \esp\
has far higher resolving power (65\,000 compared to 2\,000 or less)
than FORS1, and thus can fully exploit the Zeeman signal contained in
the metal lines. If the star is rotating slowly enough, and is cool
enough, this gives \esp\ a very large advantage over FORS1
observations, which must rely to a large extent on the signal in the
Balmer lines.

In addition, with \esp\ we have the possibility of detecting a real
and significant circular polarisation $V$ signal in the averaged line
profile even if the mean longitudinal field \bz\ is close to
zero. In weak-field stars such as HD~108945 the field is much easier
to detect this way than through conventional measurements of \bz. This
capability is also very helpful in deciding which proposed Ap stars in
a cluster are in fact Ap stars, and which are normal or belong to
other, non-magnetic, peculiarity classes. Having clear evidence on
this point is very helpful in cleaning the field measurement sample of
non-magnetic stars. 

Furthermore, with \esp\ we obtain not only spectropolarimetry, but
also high-resolution $I$ spectra over a wide wavelength window. These
data are very useful for deciding which stars have the chemical
peculiarities of magnetic Ap stars, and which appear chemically more
or less normal. We have exploited this possibility above to eliminate
some stars that have been suggested as magnetic Ap stars from the
survey sample. In a later stage of this project, we plan to model in
detail a number of the $I$ spectra from \esp\ with the goal of
obtaining a clearer picture of the evolution of atmospheric chemical
peculiarities during the main sequence lifetime of such stars. This
cannot be done at present with useful accuracy for the low resolution
spectra obtained with FORS1.

The accurate radial velocities that can be obtained with \esp\ (recall
that the precision of radial velocities, for stars with \vsi\ of the
order of 50 \kms, sometimes with non-symmetrical LSD profiles, is
roughly $\pm 1-2$ \kms) provide valuable tests of cluster membership
for stars of our sample. The accuracy of radial velocity measurements
from our FORS1 data is probably not good enough to provide really
useful constraints.

The precision of the best field measurements obtained with \esp\ is
extremely high. The best measurement have standard errors in the range
of 15--30~G, as small as or smaller than the highest precision (about
30~G) that has been obtained reliably with FORS1. This has made it
possible to add five new stars to the sample of cluster Ap stars with
field measurements, and to add further measurements of several others
stars of the sample. The new measurements provide valuable additional
data on stars having low mass and large fractional ages $\tau$, a part
of the sample which was under-represented in our previous data set.

We have re-examined the full sample, and present the data in a
somewhat different form than in Paper~II. We find that our earlier
conclusions (see Paper~II) are somewhat altered. As before, we find
that magnetic fields of stars with masses above about $3 \mo$ decline
rather strongly through the main sequence life, dropping in strength
by a factor of several after about $4\,10^7$~yr and $1.5\,10^7$~yr for
stars in the 3 -- 4 and 4 -- 5 \mo\ bins respectively. This is
consistent with the $3\,10^7$~yr found for stars above $3 \mo$ in
Paper~II. However, in contrast to the results of Paper~II, we now find
clear evidence that the fields of lower mass Ap stars (2 -- 3 \mo)
appear to decline with increasing age after roughly
$2.5\,10^8$~yr. Furthermore, for all masses, the data suggest that
even the emergent magnetic {\em flux} either declines somewhat in all
stars through the main sequence lifetime, or that the emergent flux in
the more strongly magnetic stars found at young fractional ages is
somehow reduced.

 When our results are compared with the numerical computations of
  Brathwaite \& Spruit (\cite{BraSpr04}) and Braithwaite \& Nordlund
  (\cite{BraNor06}), the data are found to support the numerical
  results in that both observations and numerical simulations agree
  that the apparently fossil fields of magnetic Ap stars are stable on
  an evolutionary time scale. On the other hand, the numerical models
  predict that during the main sequence phase the typical magnetic
  fields of Ap stars should be observed to {\it increase}, the
  opposite of what is observed. However, since these computations
  neglect both the internal (poloidal and toroidal) circulation flows
  induced by rotation (which will probably strongly distort the
  field), and the global radius increase as the star evolves from ZAMS
  to TAMS, it is quite possible that the evolutionary predictions will
  be substantially modified as more physics is included in the
  modelling. It will be of great interest to see whether including
  these (or other) physical effects in the numerical models leads to
  more complete agreement between computation and observation.

We are continuing our survey, both to obtain more precise values of
the characteristic field strength indicator \brms\ used in this study
for stars already observed, and to increase the sample to a larger
number of clusters and stars. Our ultimate goal is to provide clear
and unambiguous observational constraints on the evolution of both
magnetic field strength (and perhaps even structure), and of
atmospheric chemistry, through the main sequence lifetime of the
magnetic Ap and Bp stars. 

\acknowledgements{ 
We thank the anonymous referee for helpful comments and suggestions. 
Work by JDL, JS, JS, and GAW has been supported by the Natural
Sciences and Engineering Research Council of Canada. GAW has also been
supported by the ARP programme of DND Canada. LF has received
support from the Austrian Science Foundation (FWF project
P17980-N2). SVB acknowledges the EURYI Award from the ESF/SNF and
research grant 115417 from the Academy of Finland. This work is based
on data collected at the Canada-France-Hawaii Telescope.  Our research
has made use of the SIMBAD database, operated at CDS, Strasbourg,
France, of the WEBDA open cluster database, created at the EPFL and
now hosted at the Institute of Astronomy of the University of Vienna,
and of the Vienna Atomic Line Database VALD.  
}


\end{document}